\pdfoutput=1

\documentclass[12pt,a4paper]{article}

\usepackage{ifthen} 
\newboolean{pdflatex}
\setboolean{pdflatex}{true} 

\newboolean{articletitles}
\setboolean{articletitles}{true} 

\newboolean{uprightparticles}
\setboolean{uprightparticles}{false} 

\newboolean{inbibliography}
\setboolean{inbibliography}{false} 

\usepackage{multirow}
\usepackage{booktabs}
\usepackage{longtable}
\usepackage{amsmath}
\usepackage{verbatim}
\usepackage{epstopdf}
\usepackage{subfig}
\usepackage{rotating}
\usepackage{graphicx}

\def\paperauthors{LHCb collaboration} 
\def\paperasciititle{Precision measurement of the Bc mass} 
\def\papertitle{Precision measurement\\ of the $B_c^+$ meson mass} 
\def\paperkeywords{{High Energy Physics}, {LHCb}} 
\def\papercopyright{\the\year\ CERN for the benefit of the LHCb collaboration} 
\def\paperlicence{CC BY 4.0 licence}
\def\paperlicenceurl{https://creativecommons.org/licenses/by/4.0/}

\usepackage[top=1in, bottom=1.25in, left=1in, right=1in]{geometry}

\usepackage{microtype}
\usepackage{lineno}  
\usepackage{xspace} 
\usepackage{caption} 

\usepackage{graphicx}  
\usepackage{color}
\usepackage{colortbl}
\graphicspath{{./figs/}} 
\DeclareGraphicsExtensions{.pdf,.PDF,png,.PNG}

\usepackage{amsmath} 
\usepackage{amssymb}
\usepackage{amsfonts}
\usepackage{upgreek} 

\newcommand*\patchAmsMathEnvironmentForLineno[1]{%
\expandafter\let\csname old#1\expandafter\endcsname\csname #1\endcsname
\expandafter\let\csname oldend#1\expandafter\endcsname\csname
end#1\endcsname
 \renewenvironment{#1}%
   {\linenomath\csname old#1\endcsname}%
   {\csname oldend#1\endcsname\endlinenomath}%
}
\newcommand*\patchBothAmsMathEnvironmentsForLineno[1]{%
  \patchAmsMathEnvironmentForLineno{#1}%
  \patchAmsMathEnvironmentForLineno{#1*}%
}
\AtBeginDocument{%
\patchBothAmsMathEnvironmentsForLineno{equation}%
\patchBothAmsMathEnvironmentsForLineno{align}%
\patchBothAmsMathEnvironmentsForLineno{flalign}%
\patchBothAmsMathEnvironmentsForLineno{alignat}%
\patchBothAmsMathEnvironmentsForLineno{gather}%
\patchBothAmsMathEnvironmentsForLineno{multline}%
\patchBothAmsMathEnvironmentsForLineno{eqnarray}%
}

\usepackage{hyperxmp}

\usepackage[
            pdfauthor={\paperauthors},
            pdftitle={\paperasciititle},
            pdfkeywords={\paperkeywords},
            pdfcopyright={Copyright (C) \papercopyright},
            pdflicenseurl={\paperlicenceurl}]{hyperref}

\usepackage[colorinlistoftodos,textsize=scriptsize]{todonotes}

\usepackage[all]{hypcap} 

\usepackage{xspace} 
\usepackage{upgreek}

\def\MagUp {\mbox{\em Mag\kern -0.05em Up}\xspace}

\ifthenelse{\boolean{uprightparticles}}%
{

 \def\Ppi         {\ensuremath{\uppi}\xspace}

 \def\Ppsi        {\ensuremath{\uppsi}\xspace}

 \def\PDelta      {\ensuremath{\Delta}\xspace}                 
 \def\PXi         {\ensuremath{\Xi}\xspace}                 
 \def\PLambda     {\ensuremath{\Lambda}\xspace}                 
 \def\PSigma      {\ensuremath{\Sigma}\xspace}                 
 \def\POmega      {\ensuremath{\Omega}\xspace}                 
 \def\PUpsilon    {\ensuremath{\Upsilon}\xspace}

 \def\PB      {\ensuremath{\mathrm{B}}\xspace}                 
                  
 \def\PD      {\ensuremath{\mathrm{D}}\xspace}

 \def\PJ      {\ensuremath{\mathrm{J}}\xspace}                 
 \def\PK      {\ensuremath{\mathrm{K}}\xspace}

 \def\Pi      {\ensuremath{\mathrm{i}}\xspace}

 \def\Ps      {\ensuremath{\mathrm{s}}\xspace}

 \def\thebaroffset{0.0em}
}
{

 \def\Ppi         {\ensuremath{\pi}\xspace}

 \def\Ppsi        {\ensuremath{\psi}\xspace}                 
                  
 \mathchardef\PDelta="7101
 \mathchardef\PXi="7104
 \mathchardef\PLambda="7103
 \mathchardef\PSigma="7106
 \mathchardef\POmega="710A
 \mathchardef\PUpsilon="7107
                  
 \def\PB      {\ensuremath{B}\xspace}                 
                  
 \def\PD      {\ensuremath{D}\xspace}

 \def\PJ      {\ensuremath{J}\xspace}                 
 \def\PK      {\ensuremath{K}\xspace}

 \def\Pi      {\ensuremath{i}\xspace}

 \def\Ps      {\ensuremath{s}\xspace}

 \def\thebaroffset{0.18em}
}
\newcommand{\offsetoverline}[2][\thebaroffset]{\kern #1\overline{\kern -#1 #2}}%

\makeatletter
\ifcase \@ptsize \relax
  \newcommand{\miniscule}{\@setfontsize\miniscule{4}{5}}
\or
  \newcommand{\miniscule}{\@setfontsize\miniscule{5}{6}}
\or
  \newcommand{\miniscule}{\@setfontsize\miniscule{5}{6}}
\fi
\makeatother

\DeclareRobustCommand{\optbar}[1]{\shortstack{{\miniscule (\rule[.5ex]{1.25em}{.18mm})}
  \\ [-.7ex] $#1$}}

\def\squark    {{\ensuremath{\Ps}}\xspace}

\def\pion   {{\ensuremath{\Ppi}}\xspace}

\def\pip    {{\ensuremath{\pion^+}}\xspace}

\def\KorKbar {\kern \thebaroffset\optbar{\kern -\thebaroffset \PK}{}\xspace}

\def\DorDbar {\kern \thebaroffset\optbar{\kern -\thebaroffset \PD}\xspace}

\def\B       {{\ensuremath{\PB}}\xspace}

\def\BorBbar {\kern \thebaroffset\optbar{\kern -\thebaroffset \PB}\xspace}

\def\Bd      {{\ensuremath{\B^0}}\xspace}

\def\BdorBdbar {\kern \thebaroffset\optbar{\kern -\thebaroffset \Bd}\xspace}

\def\Bs      {{\ensuremath{\B^0_\squark}}\xspace}

\def\BsorBsbar {\kern \thebaroffset\optbar{\kern -\thebaroffset \Bs}\xspace}

\def\jpsi     {{\ensuremath{{\PJ\mskip -3mu/\mskip -2mu\Ppsi\mskip 2mu}}}\xspace}

\def\Y#1S{\ensuremath{\PUpsilon{(#1S)}}\xspace}

\def\LorLbar     {\kern \thebaroffset\optbar{\kern -\thebaroffset \PLambda}\xspace}

\def\AT#1     {\ensuremath{A_{\mathrm{T}}^{#1}}\xspace}           

\def\C#1      {\ensuremath{\mathcal{C}_{#1}}\xspace}                       
\def\Cp#1     {\ensuremath{\mathcal{C}_{#1}^{'}}\xspace}                    
\def\Ceff#1   {\ensuremath{\mathcal{C}_{#1}^{\mathrm{(eff)}}}\xspace}        
\def\Cpeff#1  {\ensuremath{\mathcal{C}_{#1}^{'\mathrm{(eff)}}}\xspace}       
\def\Ope#1    {\ensuremath{\mathcal{O}_{#1}}\xspace}                       
\def\Opep#1   {\ensuremath{\mathcal{O}_{#1}^{'}}\xspace}                    

\newcommand{\nospaceunit}[1]{\ensuremath{\text{#1}}}       
\newcommand{\aunit}[1]{\ensuremath{\text{\,#1}}}       

\newcommand{\tev}{\aunit{Te\kern -0.1em V}\xspace}
\newcommand{\gev}{\aunit{Ge\kern -0.1em V}\xspace}
\newcommand{\mev}{\aunit{Me\kern -0.1em V}\xspace}
\newcommand{\kev}{\aunit{ke\kern -0.1em V}\xspace}
\newcommand{\ev}{\aunit{e\kern -0.1em V}\xspace}
\newcommand{\mevc}{\ensuremath{\aunit{Me\kern -0.1em V\!/}c}\xspace}
\newcommand{\gevc}{\ensuremath{\aunit{Ge\kern -0.1em V\!/}c}\xspace}
\newcommand{\mevcc}{\ensuremath{\aunit{Me\kern -0.1em V\!/}c^2}\xspace}
\newcommand{\gevcc}{\ensuremath{\aunit{Ge\kern -0.1em V\!/}c^2}\xspace}

\def\gsim{{~\raise.15em\hbox{$>$}\kern-.85em
          \lower.35em\hbox{$\sim$}~}\xspace}
\def\lsim{{~\raise.15em\hbox{$<$}\kern-.85em
          \lower.35em\hbox{$\sim$}~}\xspace}

\def\tell1  {TELL1\xspace}
\def\ukl1   {UKL1\xspace}

\usepackage{cite} 
\usepackage{mciteplus}

\RequirePackage[normalem]{ulem} 
\RequirePackage{color}\definecolor{RED}{rgb}{1,0,0}\definecolor{BLUE}{rgb}{0,0,1} 

\begin{document}

\renewcommand{\thefootnote}{\fnsymbol{footnote}}
\setcounter{footnote}{1}


\begin{titlepage}
\pagenumbering{roman}

\vspace*{-1.5cm}
\centerline{\large EUROPEAN ORGANIZATION FOR NUCLEAR RESEARCH (CERN)}
\vspace*{1.5cm}
\noindent
\begin{tabular*}{\linewidth}{lc@{\extracolsep{\fill}}r@{\extracolsep{0pt}}}
\ifthenelse{\boolean{pdflatex}}
{\vspace*{-1.5cm}\mbox{\!\!\!\includegraphics[width=.14\textwidth]{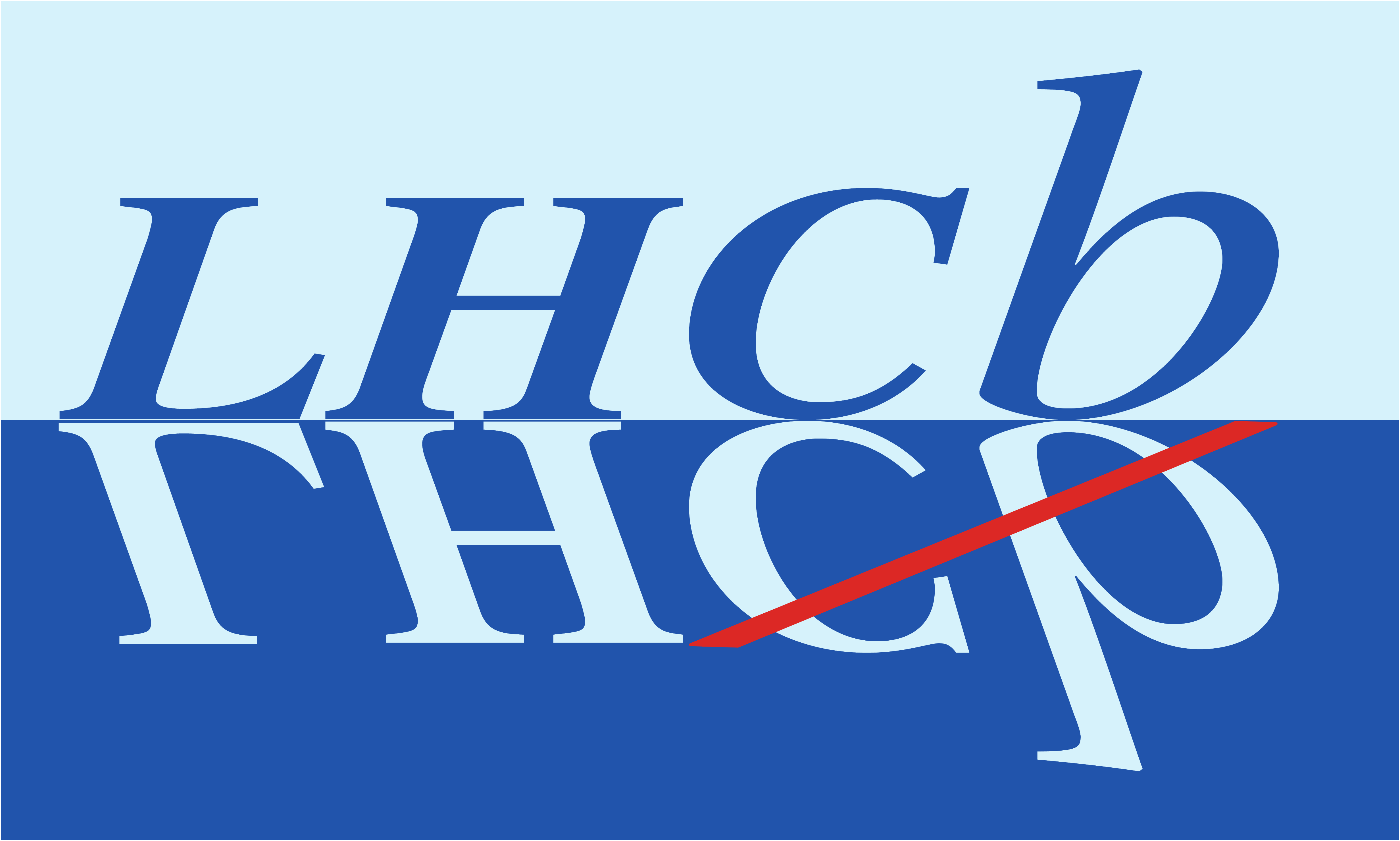}} & &}%
{\vspace*{-1.2cm}\mbox{\!\!\!\includegraphics[width=.12\textwidth]{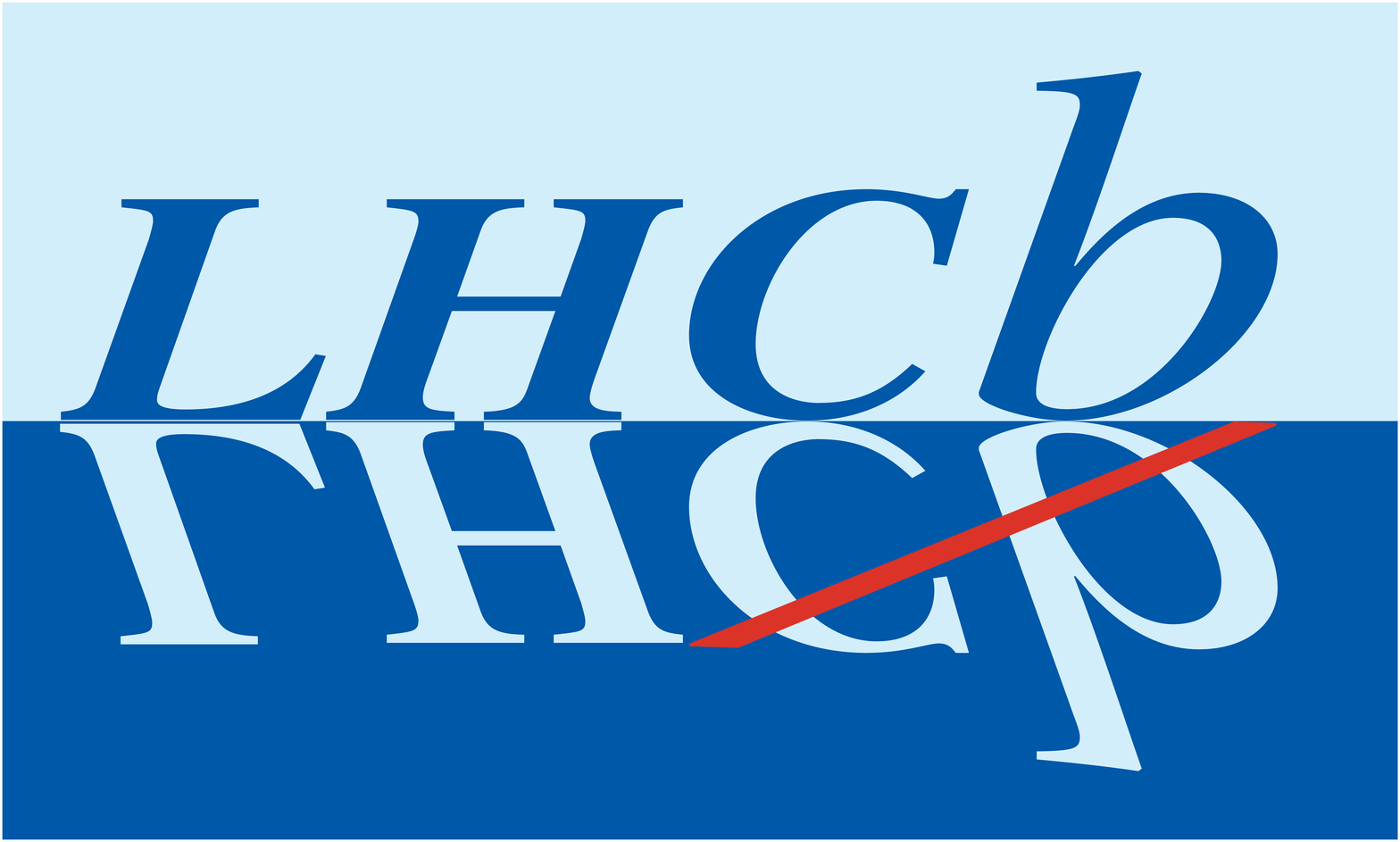}} & &}%
\\
 & & CERN-EP-2020-048 \\  
 & & LHCb-PAPER-2020-003 \\  
 & & June 3, 2020 \\ 
\end{tabular*}

\vspace*{4.0cm}

{\normalfont\bfseries\boldmath\huge
\begin{center}
  \papertitle 
\end{center}
}

\vspace*{2.0cm}

\begin{center}
\paperauthors\footnote{Authors are listed at the end of this paper.}
\end{center}

\vspace{\fill}

\begin{abstract}
  \noindent
  A precision measurement of the $B_{c}^{+}$ meson mass is performed 
  using proton-proton collision data
  collected with the LHCb experiment at centre-of-mass energies of $7, 8$
  and $13$ TeV, corresponding to a total integrated luminosity of $9.0$
  fb$^{-1}$.
  The $B_{c}^{+}$ mesons are reconstructed via the decays
  $B_{c}^{+} \rightarrow J\mskip -3mu/\mskip -2mu\psi\mskip 2mu \pi^+$,
  $B_{c}^{+} \rightarrow J\mskip -3mu/\mskip -2mu\psi\mskip 2mu \pi^+ \pi^- \pi^+$,
  $B_{c}^{+} \rightarrow J\mskip -3mu/\mskip -2mu\psi\mskip 2mu p \bar{p} \pi^+$,
  $B_{c}^{+} \rightarrow J\mskip -3mu/\mskip -2mu\psi\mskip 2mu D_{s}^{+}$, 
  $B_{c}^{+} \rightarrow J\mskip -3mu/\mskip -2mu\psi\mskip 2mu D^{0} K^{+}$
  and $B_{c}^{+} \rightarrow B_{s}^{0} \pi^+$.
  Combining the results of the individual decay channels, the $B_{c}^{+}$ mass is measured to be 
  $6274.47 \pm 0.27 \,({\rm stat}) \pm 0.17 \,({\rm syst}) \mathrm{\,Me\kern -0.1em V}/c^{2}$. 
  This is the most
  precise measurement of the $B_{c}^{+}$ mass to date. 
  The difference between the $B_{c}^{+}$ and $B_{s}^{0}$ meson masses is measured to be
  $907.75 \pm 0.37 \,({\rm stat}) \pm 0.27 \,({\rm syst}) \mathrm{\,Me\kern -0.1em V}/c^{2}$.
\end{abstract}

\vspace*{2.0cm}

\begin{center}
 Published in JHEP 07(2020) 123
\end{center}

\vspace{\fill}

{\footnotesize 
\centerline{\copyright~\papercopyright. \href{\paperlicenceurl}{\paperlicence}.}}
\vspace*{2mm}

\end{titlepage}


\newpage
\setcounter{page}{2}
\mbox{~}
%
%
%
%


\renewcommand{\thefootnote}{\arabic{footnote}}
\setcounter{footnote}{0}

\cleardoublepage

\pagestyle{plain} 
\setcounter{page}{1}
\pagenumbering{arabic}


\section{Introduction}
\label{sec:Introduction}

The $B_{c}$ meson family is unique in the Standard Model as its states
contain two different heavy-flavour quarks,
a $\bar{b}$ and a $c$ quark. 
Quantum Chromodynamics (QCD) predicts that the $\bar{b}$ and $c$ quarks are tightly bound in a compact system, with a rich spectroscopy of excited states.
Studies of the $B_{c}$ mass spectrum
can reveal information on heavy-quark dynamics
and improve our understanding of the strong interaction.
Due to the presence of two heavy-flavour quarks the mass spectrum of the $B_{c}$ states can be predicted with much better precision than many other hadronic systems.
The mass spectrum of the $B_{c}$ family has been calculated with
nonrelativistic quark potential models~\cite{Gershtein:1987jj, Chen:1992fq,
Eichten:1994gt, Kiselev:1994rc, Gupta:1995ps, Fulcher:1998ka, Devlani:2014nda, Soni:2017wvy},
nonperturbative phenomenological models~\cite{Wei:2010zza, Chen:2020ecu},
perturbative QCD~\cite{Brambilla:2000db, Xiao:2013lia},
relativistic quark models~\cite{Ebert:2002pp, Godfrey:2004ya, Ebert:2011jc, Fischer:2014cfa, Monteiro:2016rzi},
and lattice QCD~\cite{Davies:1996gi, Shanahan:1999mv, Allison:2004hy, Allison:2004be, Chiu:2007bc, Dowdall:2012ab}.
The ground state of the $B_{c}$ meson family, denoted hereafter as $B_{c}^{+}$, 
decays only through the weak interaction, with a relatively
long lifetime.
The most accurate prediction
of the $B_{c}^{+}$ mass, $M(B_{c}^{+}) = 6278 \pm 6 \pm 4 \mathrm{\,Me\kern -0.1em V}/c^{2}$~\cite{Chiu:2007bc}, is obtained with unquenched lattice QCD.

In 1998 the CDF collaboration discovered the $B_{c}^{+}$ meson via its semileptonic decay modes and measured its mass to be 
$6400 \pm 390 \pm 130 \mathrm{\,Me\kern -0.1em V}/c^{2}$~\cite{Abe:1998wi,*Abe:1998fb}.
At the LHCb experiment,  considerable progress has been made on measurements of the $B_{c}^{+}$
production~\cite{LHCb-PAPER-2012-028, LHCb-PAPER-2013-044, LHCb-PAPER-2019-033, LHCb-PAPER-2014-050, LHCb-PAPER-2016-058},
spectroscopy~\cite{LHCb-PAPER-2012-028, LHCb-PAPER-2013-010, LHCb-PAPER-2014-039, LHCb-PAPER-2016-055, LHCb-PAPER-2019-007},
lifetime~\cite{LHCb-PAPER-2013-063, LHCb-PAPER-2014-060},
and new decay modes~\cite{LHCb-PAPER-2012-054, LHCb-PAPER-2013-021, LHCb-PAPER-2013-047, LHCb-PAPER-2015-024, LHCb-PAPER-2016-020, LHCb-PAPER-2016-001, LHCb-PAPER-2016-022, LHCb-PAPER-2016-055, LHCb-PAPER-2016-058, LHCb-PAPER-2017-035, LHCb-PAPER-2017-045}. 
The world average of the $B_{c}^{+}$ mass has an uncertainty of
$0.8 \mathrm{\,Me\kern -0.1em V}/c^{2}$~\cite{PDG2018}. This is the dominant systematic uncertainty
in the recent $\B_{c}(2S)^{(*)+}$ mass measurements~\cite{Sirunyan:2019osb, LHCb-PAPER-2019-007}.

This paper presents a precision measurement of the $B_{c}^{+}$ mass
using the decay modes $B_{c}^{+} \rightarrow J\mskip -3mu/\mskip -2mu\psi\mskip 2mu \pi^+$,
  $B_{c}^{+} \rightarrow J\mskip -3mu/\mskip -2mu\psi\mskip 2mu \pi^+ \pi^- \pi^+$,
  $B_{c}^{+} \rightarrow J\mskip -3mu/\mskip -2mu\psi\mskip 2mu p \bar{p} \pi^+$,
  $B_{c}^{+} \rightarrow J\mskip -3mu/\mskip -2mu\psi\mskip 2mu D_{s}^{+}$, 
  $B_{c}^{+} \rightarrow J\mskip -3mu/\mskip -2mu\psi\mskip 2mu D^{0} K^{+}$
  and $B_{c}^{+} \rightarrow B_{s}^{0} \pi^+$~\footnote{The inclusion of charge-conjugate modes is implied throughout this paper.}.
The first two decays are chosen for their large signal yield,
while the others have a low energy release.
As the $B_{s}^{0}$ mass is known with limited precision, 
the difference between the $B_{c}^{+}$ and $B_{s}^{0}$ masses, $\Delta M = M(B_{c}^{+})-M(B_{s}^{0})$, is also measured, 
such that improvements in the $B_{s}^{+}$ mass measurement allow for a more precise $B_{c}^{+}$ mass determination.
The data sample corresponds to an integrated luminosity of $9.0$ fb$^{-1}$,
collected with the LHCb experiment in $pp$ collisions at centre-of-mass energies of $7, 8$
  and $13$~TeV.
The integrated luminosity used in this analysis is at least three times
  the one used in previous LHCb measurements~\cite{LHCb-PAPER-2012-028,
  LHCb-PAPER-2013-010, LHCb-PAPER-2014-039, LHCb-PAPER-2016-055} and the results of this paper supersede
  those earlier $B_{c}^{+}$ mass measurements.

\section{Detector and simulation}
\label{sec:Detector}

This $\mbox{LHCb}$ detector~\cite{LHCb-DP-2008-001,LHCb-DP-2014-002} 
is a single-arm forward spectrometer covering the pseudorapidity range $2<\eta <5$,
designed for the study of particles containing $b$ or $c$
quarks. The detector includes a high-precision tracking system
consisting of a silicon-strip vertex detector surrounding the $pp$
interaction region~\cite{LHCb-DP-2014-001}, a large-area silicon-strip detector located
upstream of a dipole magnet with a bending power of about
$4{\mathrm{\,Tm}}$, and three stations of silicon-strip detectors and straw
drift tubes~\cite{LHCb-DP-2013-003,LHCb-DP-2017-001} placed downstream of the magnet.
The tracking system provides a measurement of the momentum, $p$, of charged particles with
a relative uncertainty that varies from $0.5\%$ at low momentum to $1.0\%$
at $200\mathrm{\,Ge\kern -0.1em V}/c$.
The momentum scale is calibrated using samples of $B^{+} \rightarrow J\mskip -3mu/\mskip -2mu\psi\mskip 2mu K^+$ 
and $J\mskip -3mu/\mskip -2mu\psi\mskip 2mu \rightarrow \mu^+ \mu^-$ decays collected concurrently
with the data sample used for this analysis~\cite{LHCb-PAPER-2012-048,LHCb-PAPER-2013-011}. 
The relative accuracy of 
this
procedure is determined to be $3 \times 10^{-4}$ using samples of other
fully reconstructed $B$, $\varUpsilon$ and
$K_{S}^{0}$-meson decays.
The minimum distance of a track to a primary vertex (PV), the impact parameter (IP), 
is measured with a resolution of $(15+29/p_{\mathrm{T}})\,{\mathrm{\upmu\nospaceunit{m}}}$,
where $p_{\mathrm{T}}$ is the component of the momentum transverse to the beam, in$\mathrm{\,Ge\kern -0.1em V}/c$.
Different types of charged hadrons are distinguished using information
from two ring-imaging Cherenkov detectors~\cite{LHCb-DP-2012-003}. 
Photons, electrons and hadrons are identified by a calorimeter system consisting of
a scintillating-pad and preshower detectors, an electromagnetic
and a hadronic calorimeter. Muons are identified by a
system composed of alternating layers of iron and multiwire
proportional chambers~\cite{LHCb-DP-2012-002}.
The online event selection is performed by a trigger~\cite{LHCb-DP-2012-004}, 
which consists of a hardware stage, based on information from the calorimeter and muon
systems, followed by a software stage, which performs a full event
reconstruction.

Simulated samples are used to model the effects of the detector
acceptance, optimise signal selection and
validate the analysis technique. 
In simulation, $pp$ collisions
are generated using $\textsc{Pythia}$ 8~\cite{Sjostrand:2007gs} with an $\mbox{LHCb}$
specific configuration~\cite{LHCb-PROC-2010-056}.
The production of $B_{c}^{+}$ mesons is simulated using the dedicated generator $\textsc{BcVegPy}$~\cite{Chang:2005hq}. 
Decays of hadrons are described by $\textsc{EvtGen}$~\cite{Lange:2001uf}, in which final-state radiation is generated using $\mbox{\textsc{Photos} 3}$~\cite{Golonka:2005pn}.
The interaction of the generated particles with the detector and its
response are implemented using the $\textsc{Geant4}$
toolkit~\cite{Allison:2006ve, *Agostinelli:2002hh} as described in
Ref.~\cite{LHCb-PROC-2011-006}.

\section{Event selection}
\label{sec:Eventselection}

The $B_{c}^{+}$ candidates are reconstructed in 
the following decay modes:
$B_{c}^{+} \rightarrow J\mskip -3mu/\mskip -2mu\psi\mskip 2mu \pi^+$,
$B_{c}^{+} \rightarrow J\mskip -3mu/\mskip -2mu\psi\mskip 2mu \pi^+ \pi^- \pi^+$,
$B_{c}^{+} \rightarrow J\mskip -3mu/\mskip -2mu\psi\mskip 2mu p \bar{p} \pi^+$,
$B_{c}^{+} \rightarrow J\mskip -3mu/\mskip -2mu\psi\mskip 2mu D_{s}^{+}$, 
$B_{c}^{+} \rightarrow J\mskip -3mu/\mskip -2mu\psi\mskip 2mu D^{0} K^{+}$ 
and $B_{c}^{+} \rightarrow B_{s}^{0} \pi^+$.
A pair of oppositely
charged muons form $J\mskip -3mu/\mskip -2mu\psi\mskip 2mu$ candidates. 
The $D_{s}^{+}$ candidates are reconstructed via the $D_{s}^{+} \rightarrow K^{+}
K^{-} \pi^{+}$ and $D_{s}^{+} \rightarrow \pi^{+} \pi^{-} \pi^{+}$ decays, 
while the $D^{0}$ is reconstructed
using the $D^{0} \rightarrow K^{-} \pi^{+}$ decay.
The $B_{s}^{0}$ candidates are reconstructed in the decay modes 
$B_{s}^{0} \rightarrow J\mskip -3mu/\mskip -2mu\psi\mskip 2mu(\rightarrow \mu^{+} \mu^{-}) \phi(\rightarrow K^+ K^-)$
and 
$B_{s}^{0} \rightarrow D_{s}^{-}(\rightarrow K^{+} K^{-} \pi^{-}) \pi^{+}$, 
and a multivariate classifier as used in Ref.~\cite{LHCb-PAPER-2013-044} is employed to separate signal from combinatorial background. 
Then the $B_{s}^{0}$ candidates are combined with an additional pion to reconstruct $B_{c}^{+}$
candidates. 
All of the intermediate-state particles are required to have an invariant mass
within three times the expected mass resolution around their known masses~\cite{PDG2018}. 
Muons, kaons, pions and protons
 are required to have good
 track-fit quality
 and high transverse momentum. 
The $J\mskip -3mu/\mskip -2mu\psi\mskip 2mu$ and $B_{c}^{+}$ candidates are required to have a good-quality vertex fit.

A boosted decision tree~\cite{Breiman:1984jka, Freund:1997xna, friedman2001}  implemented within the TMVA~\cite{TMVA4} package
optimises separation of the signal from combinatorial background for each decay mode.
The classifiers are trained with simulated signal samples
and a background proxy obtained from the upper mass sideband of the data, in the range $[6.6, 7.0]\mathrm{\,Ge\kern -0.1em V}/c^{2}$.
Kinematic variables that generically separate $b$-hadron decays from background are used in the training of the classifiers. The variables include the 
decay time, transverse momenta,
vertex-fit quality of the $B_{c}^{+}$ candidate, as well as variables related to the fact that the $B_{c}^{+}$ meson is produced at the PV.
The requirement on the classifiers is determined by maximising the signal
significance $S/\sqrt{S+B}$, where $S$ is the expected signal yield
estimated using simulation, and $B$ is the expected background yield
evaluated in the upper sideband in data and extrapolated to the signal
region.

\section{Mass measurement}
\label{sec:massfit}

The $B_{c}^{+}$ meson mass is determined in each decay mode by performing
an unbinned maximum likelihood fit to the invariant mass distributions of  
the $B_{c}^{+}$ candidates. The signal is described by a double-sided Crystal
Ball (DSCB) function~\cite{Skwarnicki:1986xj}, while the background is described by an exponential function.
The DSCB function comprises a Gaussian core with  power-law tails to account for 
radiative effects.
Parameters describing the radiative tails are determined from simulation.

The invariant mass of the $B_{c}^{+}$ candidates is calculated from a 
kinematic fit~\cite{Hulsbergen:2005pu},
in which the $B_{c}^{+}$ candidate is assumed to originate from its PV and the intermediate-state masses 
are constrained to their known values~\cite{PDG2018}. 
The PV of the $B_{c}^{+}$ candidate is that with respect to which it has the smallest $\chi^{2}_{\text{IP}}$. The $\chi^{2}_{\text{IP}}$ is defined as the difference in $\chi^{2}$ of the PV fit with and without the particle in question.
For $B_{c}^{+} \rightarrow B_{s}^{0} \pi^{+}$ decays, the $B_{s}^{0}$ mass is constrained to the value of  
$5366.89 \pm 0.21\mathrm{\,Me\kern -0.1em V}/c^{2}$,
which is an average of the measurements of the $B_{s}^{0}$ mass performed by the LHCb collaboration~\cite{LHCb-PAPER-2018-046, LHCb-PAPER-2018-018, LHCb-PAPER-2015-033, LHCb-PAPER-2011-035}.

The difference between the $B_{c}^{+}$ and $B_{s}^{0}$ meson masses, $\Delta m = m(B_{c}^{+})-m(B_{s}^{0})$, is determined in 
the $B_{c}^{+} \rightarrow B_{s}^{0} \pi^{+}$ decay mode, 
where $m(B_{c}^{+})$ and $m(B_{s}^{0})$ are the reconstructed masses of $B_{c}^{+}$ and $B_{s}^{0}$ candidates.
The mass difference $\Delta m$
is calculated with a 
kinematic fit~\cite{Hulsbergen:2005pu},
in which the $B_{c}^{+}$ candidate is assumed to originate from the PV with the smallest $\chi^{2}_{\text{IP}}$ and the masses of the intermediate particles 
are constrained to their known values~\cite{PDG2018}. 
The fitting procedure for the mass difference is the same as for the mass fit.

\begin{figure}
  \centering
    \resizebox{0.48\textwidth}{!}{
      \includegraphics{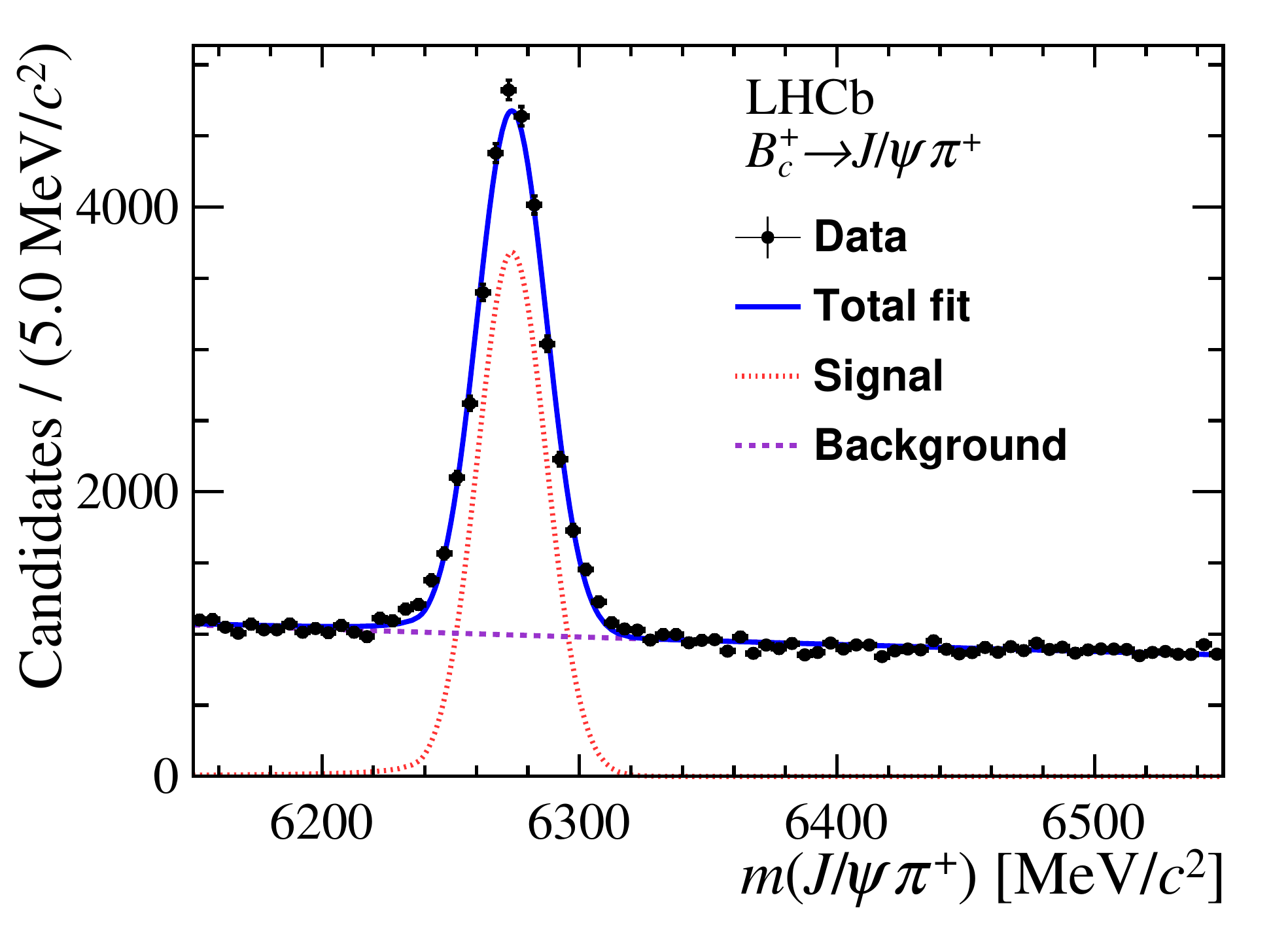}}
    \resizebox{0.48\textwidth}{!}{
      \includegraphics{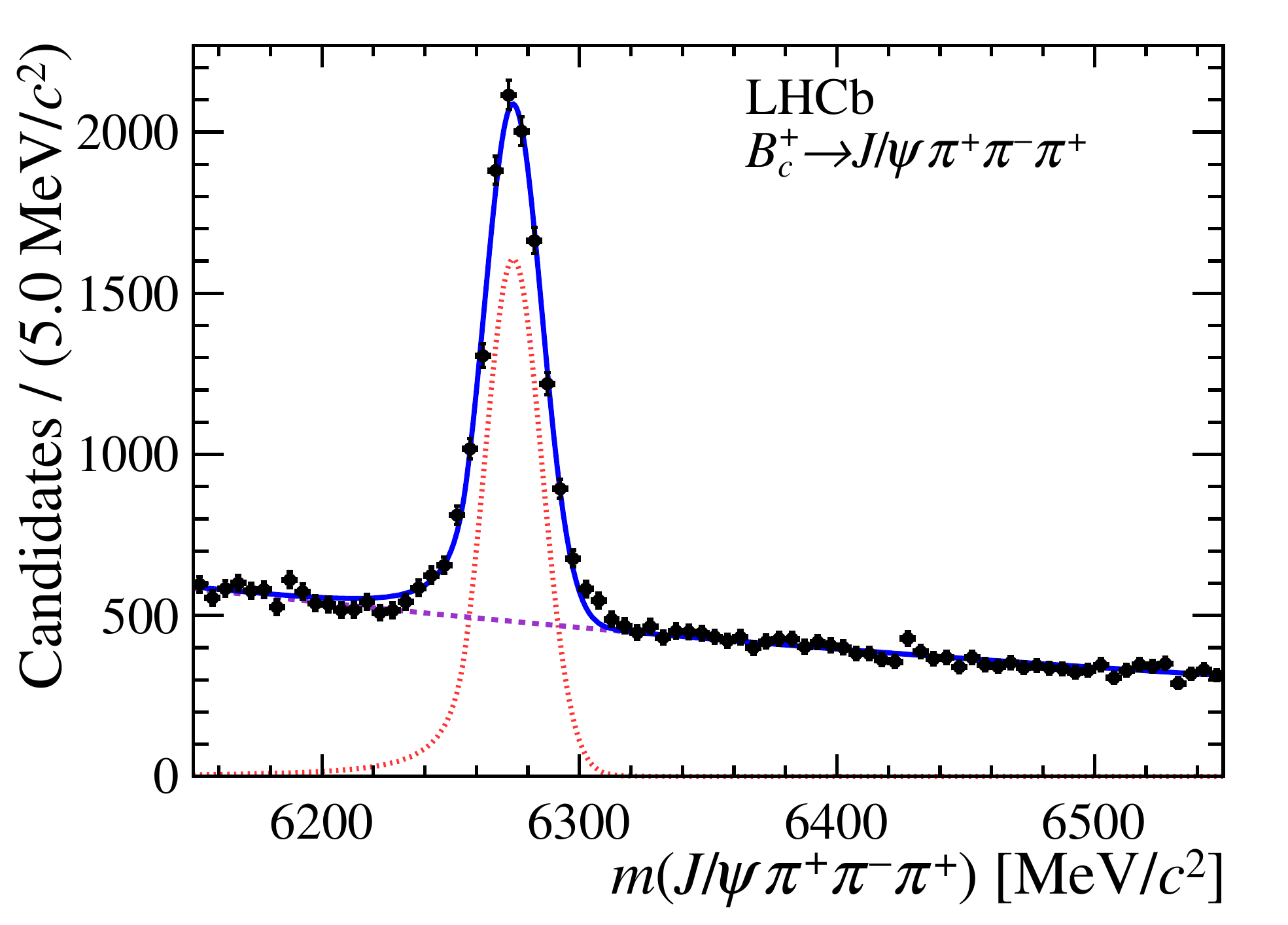}}
   \\
    \resizebox{0.48\textwidth}{!}{
      \includegraphics{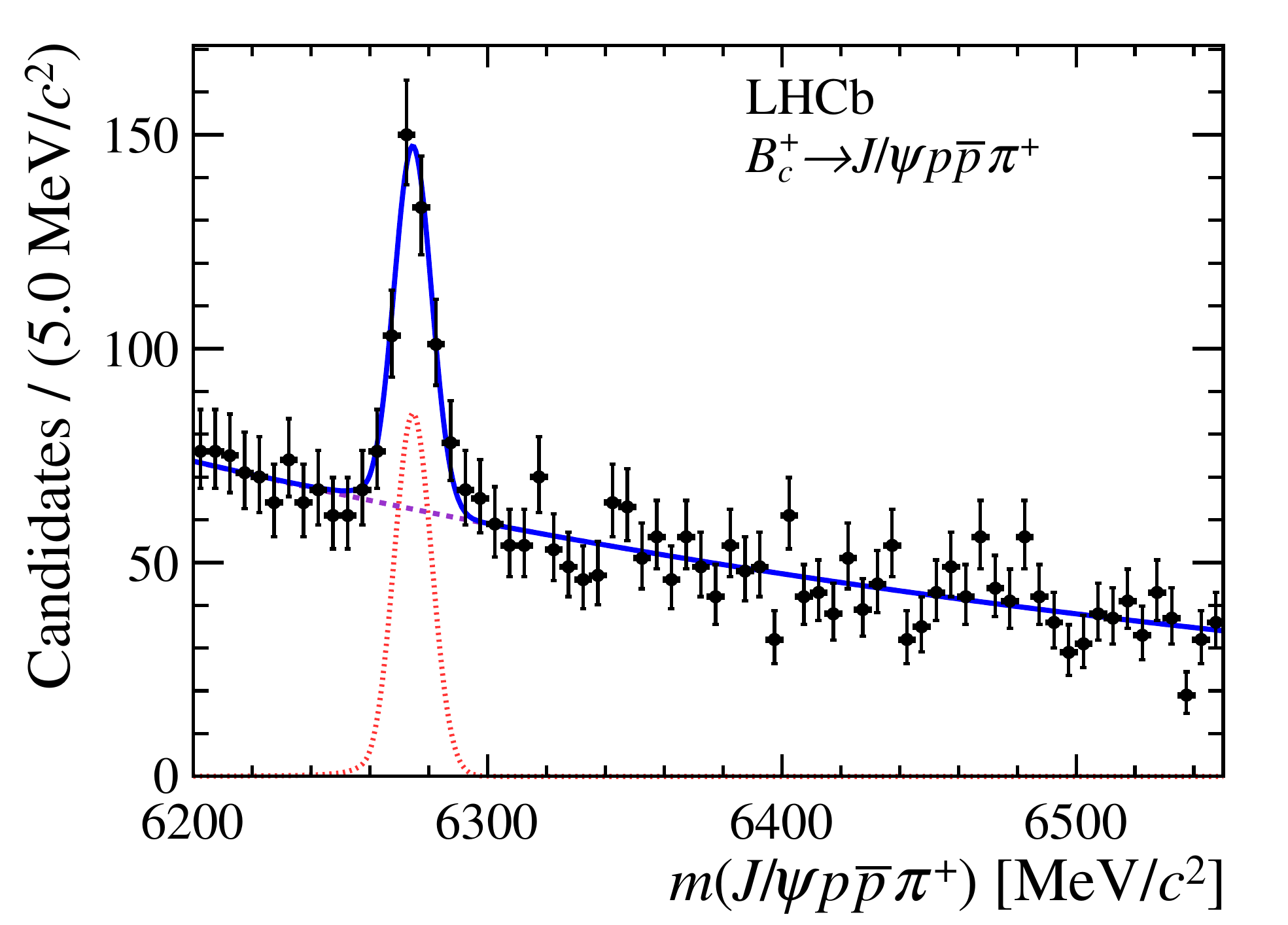}}
    \resizebox{0.48\textwidth}{!}{
      \includegraphics{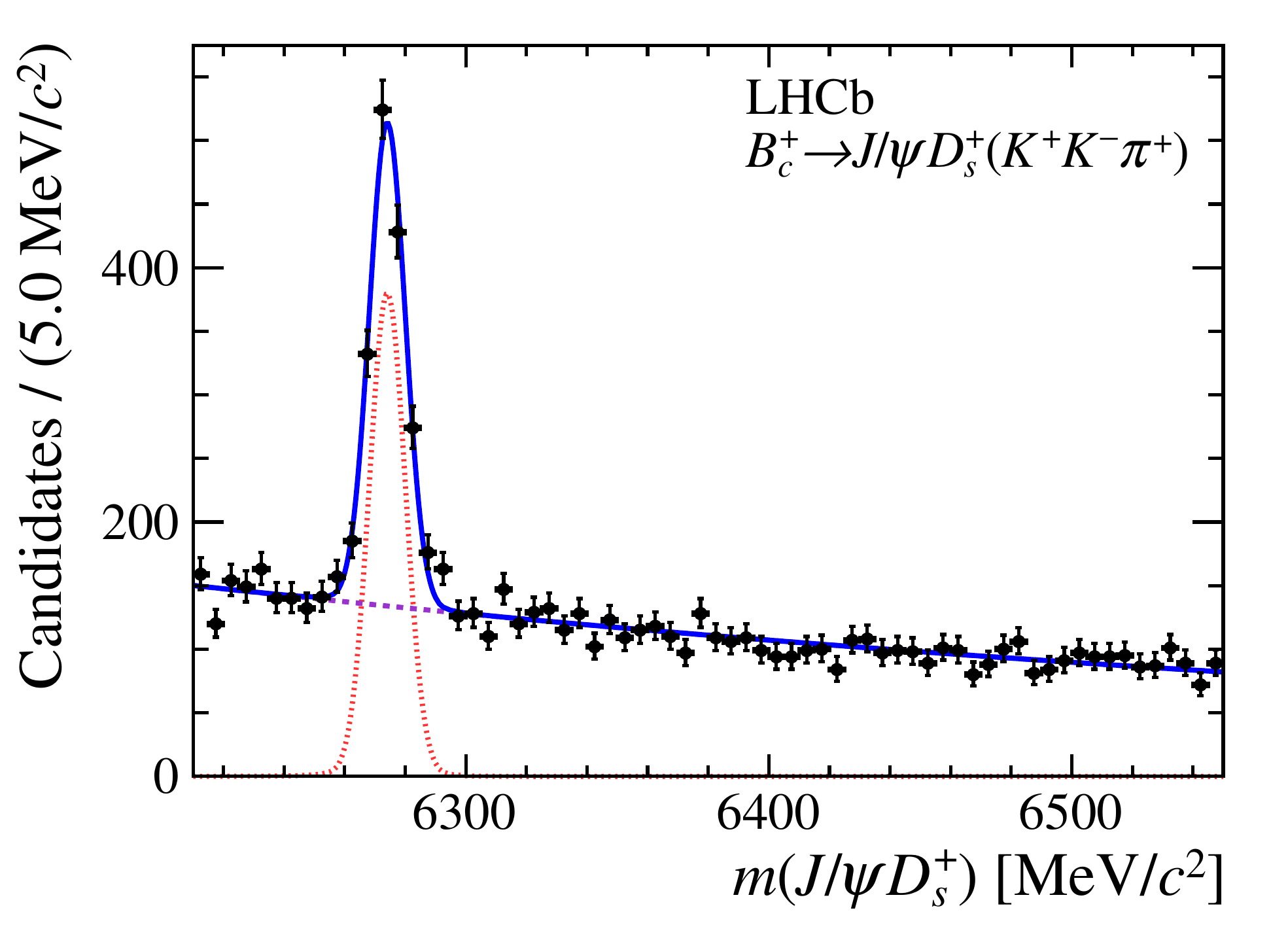}}
\\
    \resizebox{0.48\textwidth}{!}{
      \includegraphics{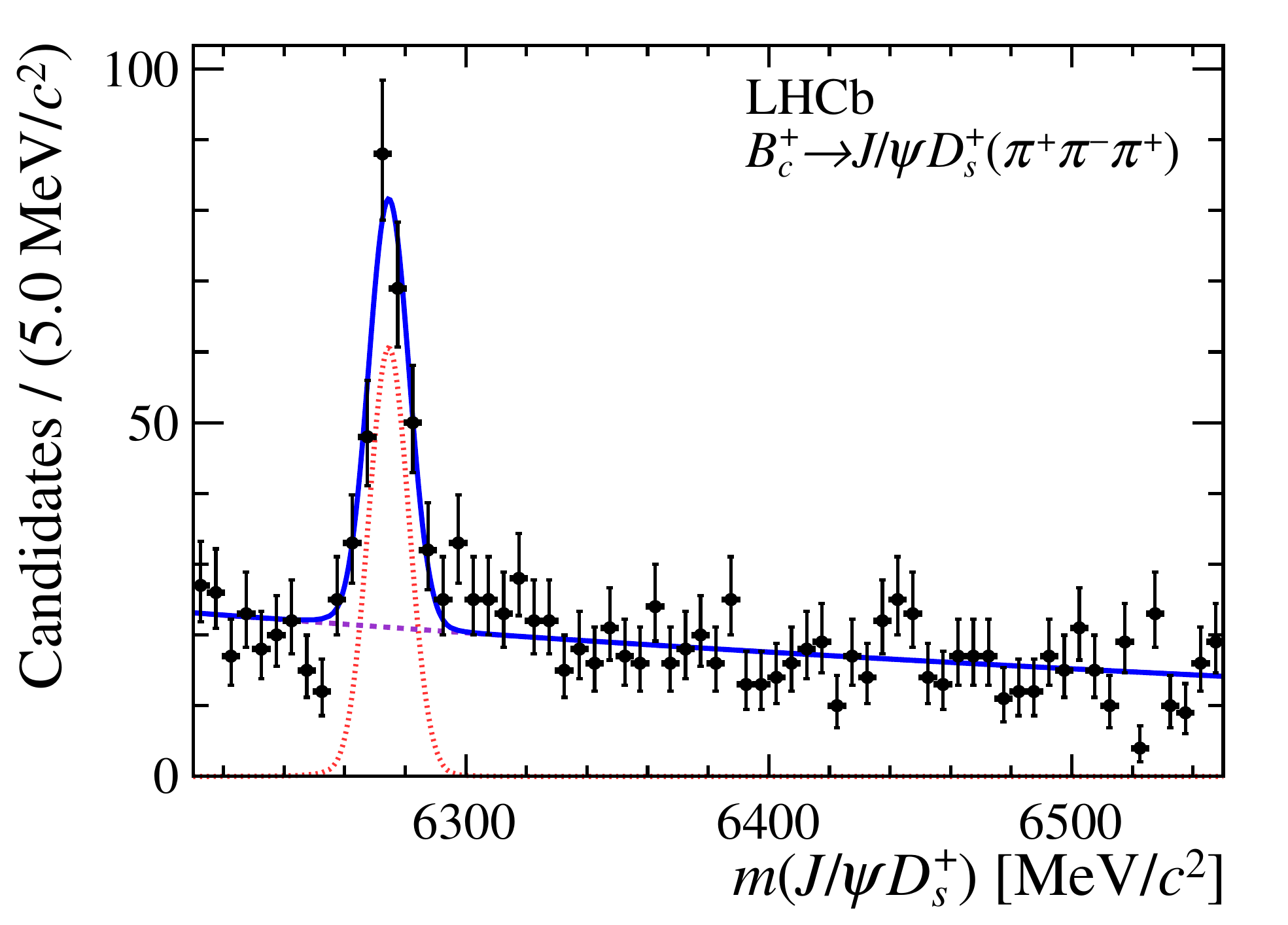}}
    \resizebox{0.48\textwidth}{!}{
      \includegraphics{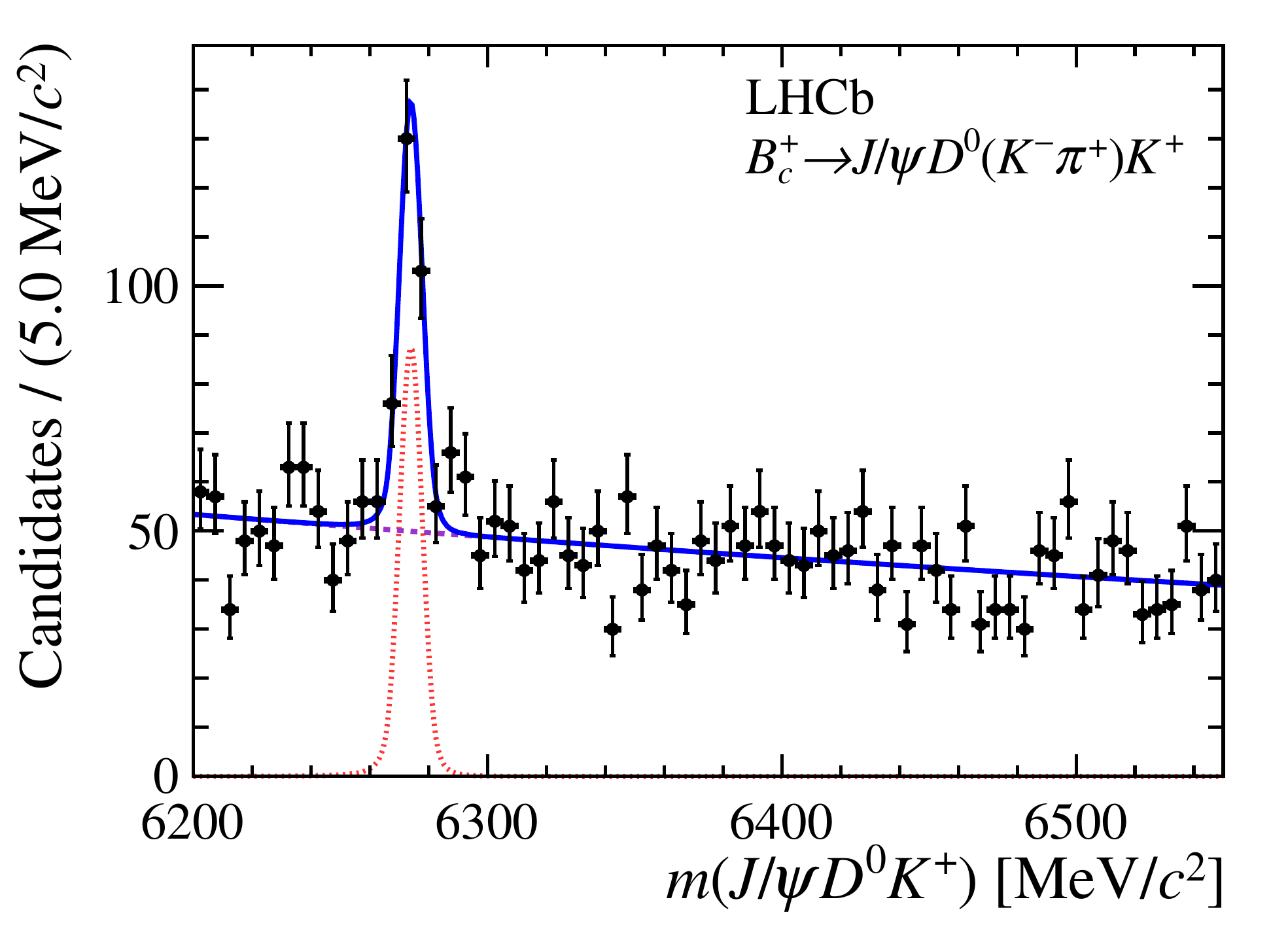}}
\\
    \resizebox{0.48\textwidth}{!}{
      \includegraphics{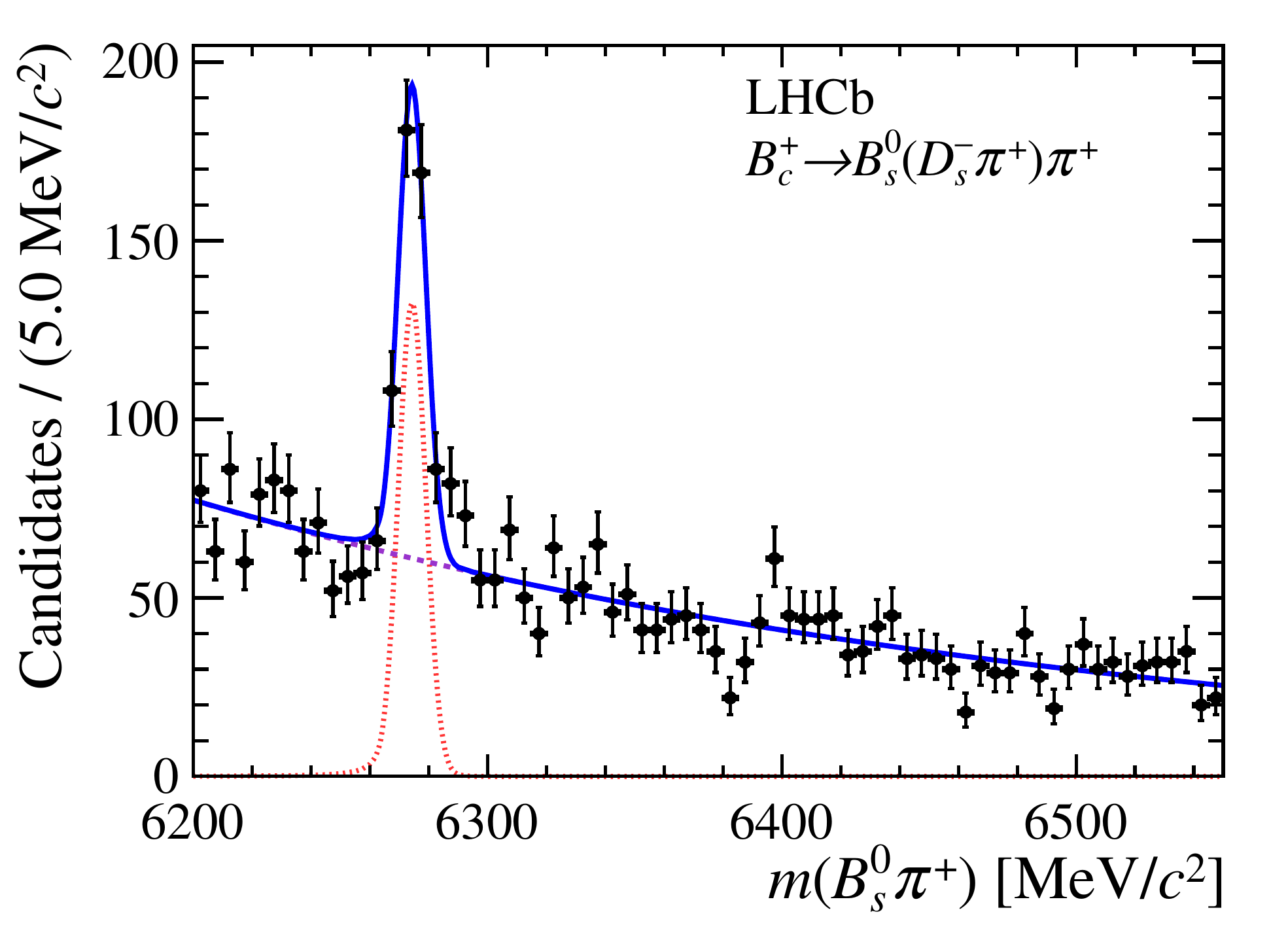}}
    \resizebox{0.48\textwidth}{!}{
      \includegraphics{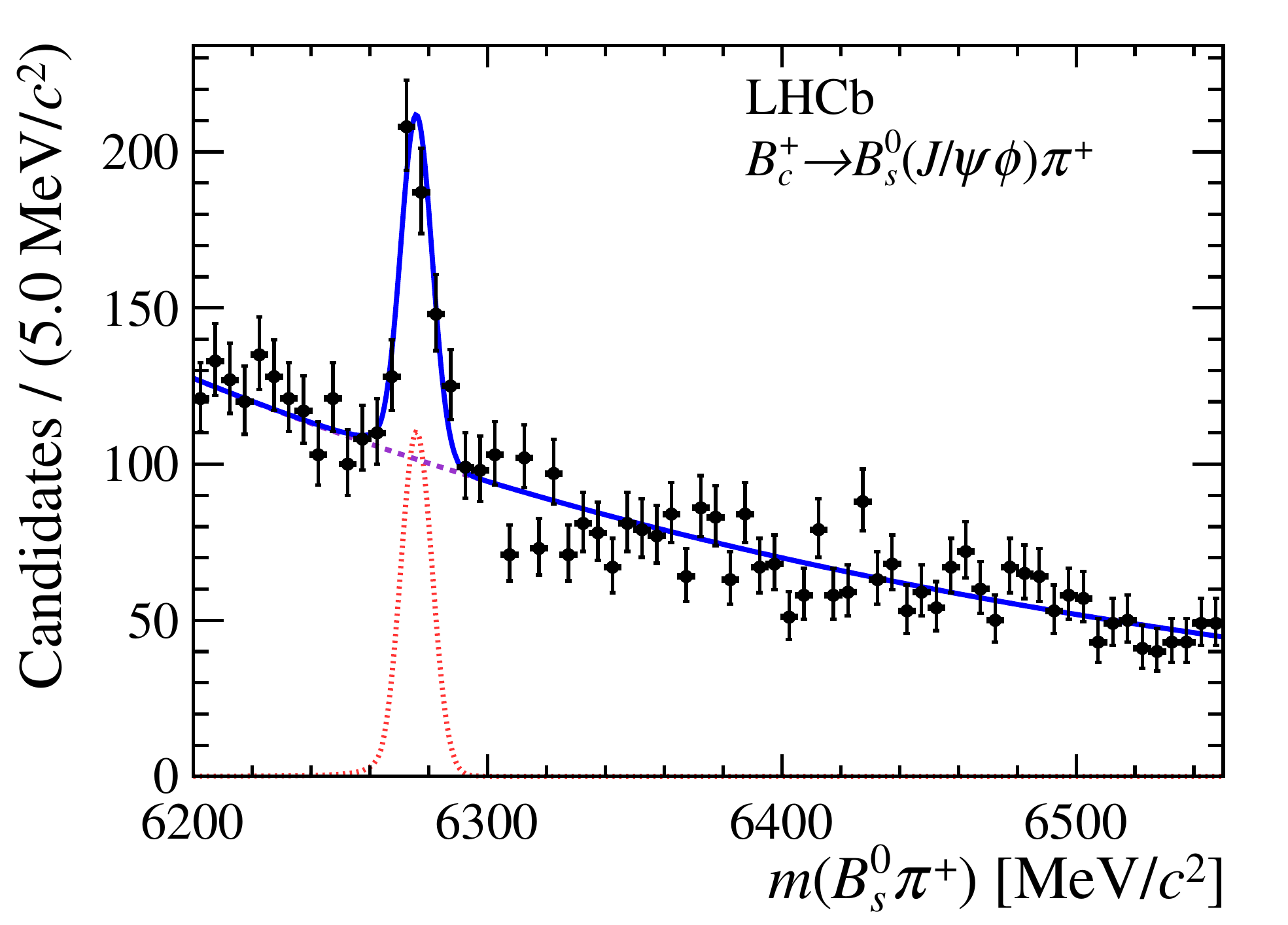}}
  \caption{Distributions of invariant-mass $m$ for $B_{c}^{+}$ candidates selected in the studied decay channels, where data are shown as the points with error bars; the total fits are shown as solid blue curves; the signal component are red dotted curves; the background components purple dotted curves.}
  \label{fig:massfit}
\end{figure}

\begin{figure}
  \centering
    \resizebox{0.48\textwidth}{!}{
      \includegraphics{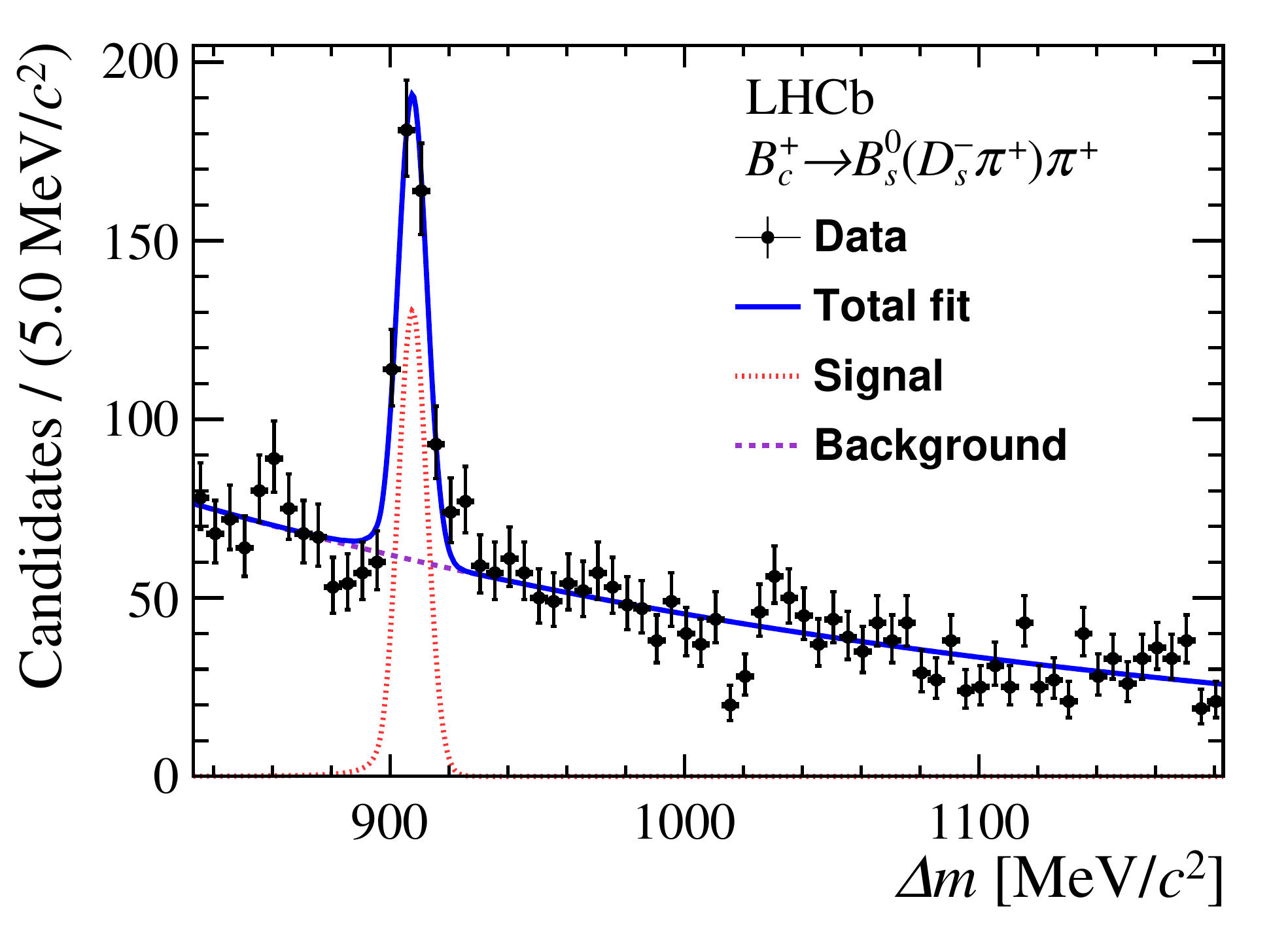}}
    \resizebox{0.48\textwidth}{!}{
      \includegraphics{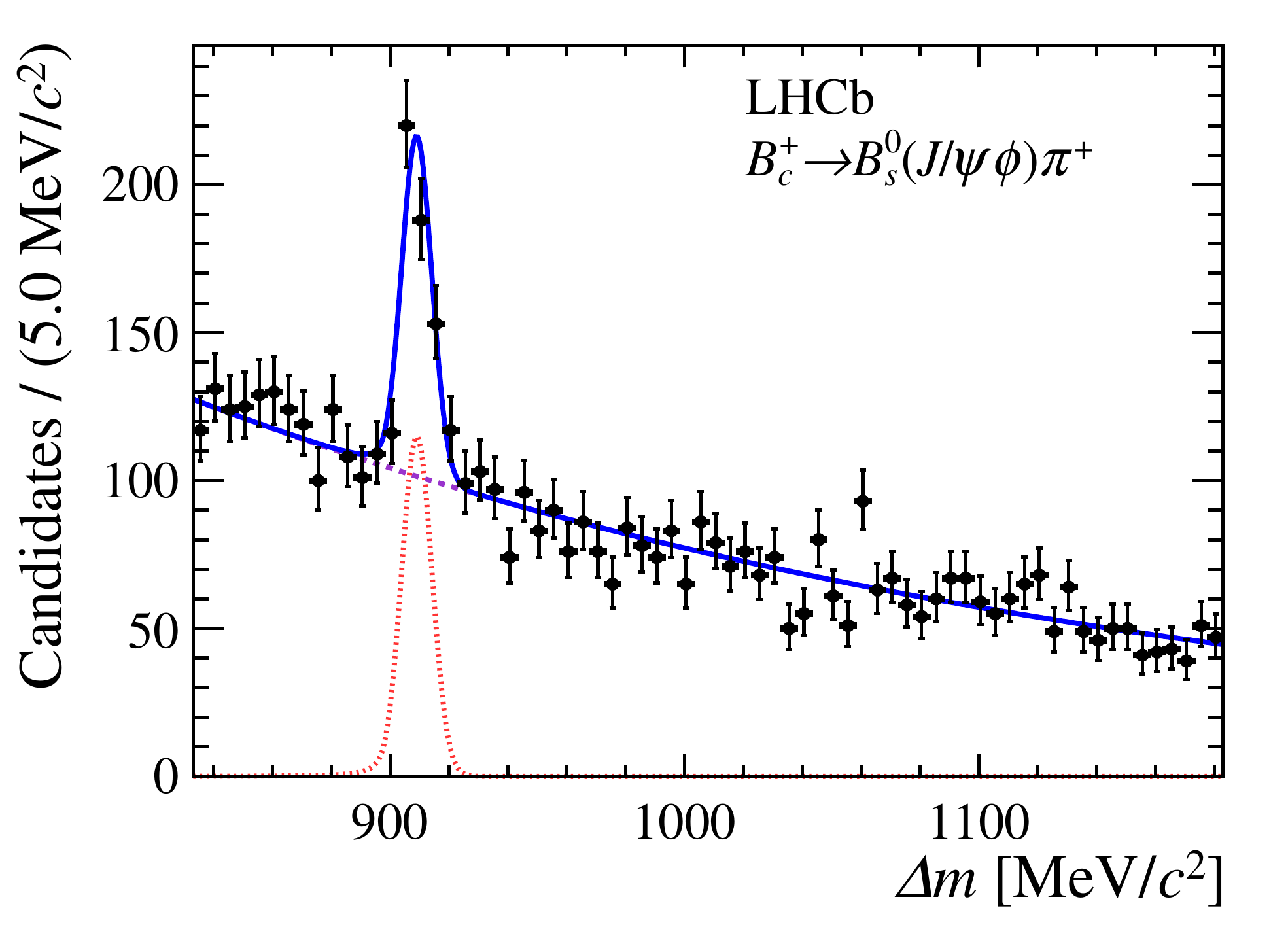}}
  \caption{Distributions of mass difference $\Delta m$ for the $B_{c}^{+} \rightarrow B_{s}^{0} (D_{s}^{-} \pi^{+}) \pi^{+}$ and $B_{c}^{+} \rightarrow B_{s}^{0} (J\mskip -3mu/\mskip -2mu\psi\mskip 2mu \phi) \pi^{+}$ decay modes, where data are shown as the points with error bars; the total fits are shown as solid blue curves; the signal component are red dotted curves; the background components purple dotted curves.}
  \label{fig:Bsmassdiff}
\end{figure}

Figure~\ref{fig:massfit} shows the invariant mass distributions and
fit results for all $B_{c}^{+}$
decay modes. Figure~\ref{fig:Bsmassdiff} shows the 
distributions of $\Delta m$
and fit results for the
$B_{c}^{+} \rightarrow B_{s}^{0} (D_{s}^{-} \pi^{+}) \pi^{+}$ and $B_{c}^{+} \rightarrow B_{s}^{0} (\jpsi \phi) \pip$ 
decay modes.
The lower limit of the mass window is chosen to exclude the partially reconstructed background while keeping sufficient left mass sideband.
The signal yields, mass and resolution values as determined from fits to the individual mass distributions 
are given in Table~\ref{tab:fitinfo}.
For the $B_{c}^{+} \rightarrow B_{s}^{0} \pi^{+}$ decays, the results of the fits to the $\Delta m$ distribution are reported in Table~\ref{tab:fitinfo_massdiff}.

The reconstructed invariant-mass distribution is distorted due to the
missing energy from unreconstructed photons (bremsstrahlung) emitted by final-state particles.
The resulting bias in the extracted $B_{c}^{+}$ mass is studied with simulated samples for each decay channel,
and is used to correct the mass obtained from the fit.
Multiple scattering in detector material can decrease the observed opening angles among the
$B_{c}^{+}$ decay products, 
affecting the reconstructed $B_{c}^{+}$ mass and
decay length and thereby the selection efficiency. 
Such effect distorts the mass distribution after the event selection.
The corresponding bias of the $B_{c}^{+}$ mass measurement
was studied with charmed hadrons ($D^{+}, D^{0}, D_{s}^{+}, \varLambda_{c}^{+}$),
and was found to be well reproduced by simulation~\cite{LHCb-PAPER-2017-018}. 
A bias associated with the selection from simulated samples is assigned as a corresponding correction.
The measured masses ($M$) and mass difference ($\Delta M$) are corrected for 
this bias (from -0.46 to 0.27\mevcc) due to final-state radiation and
the selection, and summarised in Table~\ref{tab:fitinfo} and~\ref{tab:fitinfo_massdiff}.

\begin{table}
\begin{center}
  \caption{\small \label{tab:fitinfo}
    Signal yields, mass values and mass resolutions as obtained from
    fits shown in Fig.~\ref{fig:massfit}, together with the mass corrected for the
    effects of final-state radiation and selection as described in the text. The
    uncertainties are statistical only.}
\begin{tabular}{@{}l|cccc@{}}
\hline
 \multirow{2}{*}{Decay mode} &\multirow{2}{*}{Yield} &  Fitted mass & Corrected mass & Resolution\\
                                  &                          &    [$\mathrm{\,Me\kern -0.1em V}/c^{2}\,$] & [$\mathrm{\,Me\kern -0.1em V}/c^{2}\,$] & [$\mathrm{\,Me\kern -0.1em V}/c^{2}\,$] \\
\hline
 $J\mskip -3mu/\mskip -2mu\psi\mskip 2mu \pi^+$                   & $25181 \pm 217$       & 6273.71 $\pm$ 0.12 & 6273.78 $\pm$ 0.12 & $13.49 \pm 0.11$ \\
 $J\mskip -3mu/\mskip -2mu\psi\mskip 2mu \pi^+ \pi^- \pi^+$             & $\,~9497 \pm 142$     & 6274.26 $\pm$ 0.18 & 6274.38 $\pm$ 0.18 & $11.13 \pm 0.18$ \\
 $J\mskip -3mu/\mskip -2mu\psi\mskip 2mu p \bar{p} \pi^+$          & $\,~\,~273 \pm \,~29$ & 6274.66 $\pm$ 0.73 & 6274.61 $\pm$ 0.73 & $\,~6.34 \pm 0.76$ \\
 $J\mskip -3mu/\mskip -2mu\psi\mskip 2mu D_{s}^{+} (K^+ K^- \pi^+)$     & $\,~1135 \pm \,~49$   & 6274.09 $\pm$ 0.27 & 6274.11 $\pm$ 0.27 & $\,~5.93 \pm 0.30$ \\ 
 $J\mskip -3mu/\mskip -2mu\psi\mskip 2mu D_{s}^{+} (\pi^+ \pi^- \pi^+)$ & $\,~\,~202 \pm \,~20$ & 6274.57 $\pm$ 0.71 & 6274.29 $\pm$ 0.71 & $\,~6.63 \pm 0.67$ \\ 
 $J\mskip -3mu/\mskip -2mu\psi\mskip 2mu D^{0} (K^- \pi^+) K^{+}$       & $\,~\,~175 \pm \,~21$ & 6273.97 $\pm$ 0.53 & 6274.08 $\pm$ 0.53 & $\,~3.87 \pm 0.57$ \\ 
 $B_{s}^{0} (D_s^- \pi^+) \pi^+$         & $\,~\,~316 \pm \,~27$ & 6274.36 $\pm$ 0.44 & 6274.08 $\pm$ 0.44 & $\,~4.67 \pm 0.48$ \\ 
 $B_{s}^{0} (J\mskip -3mu/\mskip -2mu\psi\mskip 2mu \phi) \pi^{+}$          & $\,~\,~299 \pm \,~37$ & 6275.87 $\pm$ 0.66 & 6275.46 $\pm$ 0.66 & $\,~5.32 \pm 0.74$ \\ 
\hline
\end{tabular}
\end{center}
\end{table}

\begin{table}
\begin{center}
  \caption{\small \label{tab:fitinfo_massdiff}
    Signal yields, mass difference $(\Delta M)$ and resolution as obtained from
    fits shown in Fig.~\ref{fig:Bsmassdiff}, together with the values corrected for the
    effects of final-state radiation and selection as described in the text. The
     uncertainties are statistical only.}
\begin{tabular}{@{}l|cccc@{}}
\hline
 \multirow{2}{*}{Decay mode} &\multirow{2}{*}{Yield} &  Fitted $\Delta M$ & Corrected $\Delta M$ & Resolution\\
                                  &                          &    [$\mathrm{\,Me\kern -0.1em V}/c^{2}\,$] & [$\mathrm{\,Me\kern -0.1em V}/c^{2}\,$] & [$\mathrm{\,Me\kern -0.1em V}/c^{2}\,$] \\
\hline
 $B_{s}^{0} (D_s^- \pi^+) \pi^+$         & $325 \pm 27$ & 907.51 $\pm$ 0.46 & 907.24 $\pm$ 0.46 & $\,~4.88 \pm 0.47$ \\ 
 $B_{s}^{0} (J\mskip -3mu/\mskip -2mu\psi\mskip 2mu \phi) \pi^{+}$          & $300 \pm 32$ & 908.98 $\pm$ 0.61 & 908.59 $\pm$ 0.61 & $\,~5.12 \pm 0.62$ \\ 
\hline
\end{tabular}
\end{center}
\end{table}

\section{Systematic uncertainties}
\label{sec:systematics}

To evaluate systematic uncertainties, the complete analysis
is repeated varying assumed parameters, models and selection requirements.
The observed differences in the 
$B_{c}^{+}$ mass central values between
the nominal result and the alternative estimates are considered as one standard-deviation uncertainties.

The systematic uncertainty of the $B_{c}^{+}$ mass comprises
uncertainties on the 
momentum-scale calibration, energy loss corrections,
signal and background models, the mass of the intermediate
states and the uncertainty on the bias caused by the final-state radiation and selection.

The dominant source of systematic uncertainty arises due to the limited precision of the momentum-scale calibration. 
For each decay, this uncertainty 
is propagated to the $B_{c}^{+}$ mass according to the energy release,
which is the difference between the value of the $B_{c}^{+}$ mass and the sum of the masses of its intermediate states. 
The amount of material traversed in the tracking system by a particle is known to $10\%$ accuracy, 
which leads to an uncertainty on the estimated energy loss. 
This translates into a measured mass uncertainty of $0.03\mathrm{\,Me\kern -0.1em V}/c^{2}$ for 
$D^{0} \rightarrow K^+ K^- \pi^+ \pi^-$ decays~\cite{LHCb-PAPER-2013-011}.
The uncertainties on the $B_{c}^{+}$ mass are scaled from that of the
$D^{0}$ decay by the number of final-state particles.
The uncertainties due to the limited size of simulated samples
are taken as systematic uncertainties from the selection-induced
bias on the $B_{c}^{+}$ masses.
The uncertainty on the masses of the intermediate states $D_{s}^{+}, D^{0}, B_{s}^{0}$ are propagated to the 
$B_{c}^{+}$ mass measurement.

The uncertainty related to the signal shape is estimated by
using alternative signal models, including
the sum of two Gaussian functions, a Hypatia function~\cite{Santos:2013gra}, 
the sum of a DSCB and a Gaussian function, 
and the sum of two DSCB functions.
The differences
of the fitted mass with final-state radiation corrections between the
nominal and the alternative models are found to be smaller than $0.1\mathrm{\,Me\kern -0.1em V}/c^{2}$, 
which is taken as the corresponding systematic uncertainty.
The uncertainty related to the background description is evaluated by
using a first-order Chebyshev function instead of an  exponential function.

The non-resonant contribution, 
for example the contribution of $B_{c}^{+} \rightarrow J\mskip -3mu/\mskip -2mu\psi\mskip 2mu \pi^+ \pi^- \pi^+$ decays to 
the $B_{c}^{+} \rightarrow J\mskip -3mu/\mskip -2mu\psi\mskip 2mu D_{s}^{+} (\pi^+ \pi^- \pi^+)$ candidates, 
is found to be highly suppressed and have negligible effects on the mass measurement.
The systematic uncertainties considered for the $B_{c}^{+}$ mass and mass difference measurements are summarised in Table~\ref{tab:systematics} 
and~\ref{tab:Bs_systematics}, respectively.

\begin{table}[bp]
\begin{center}
  \caption{\small \label{tab:systematics} Summary of systematic uncertainties (in $\mathrm{\,Me\kern -0.1em V}/c^{2}$) on the $B_{c}^{+}$ mass.}
\resizebox{\textwidth}{!}{
\begin{tabular}{@{}l|cccccc|c@{}}
\hline
  & Momentum & Energy & \multirow{2}{*}{Signal} & \multirow{2}{*}{Background} & \multirow{2}{*}{Intermediate} & \multirow{3}{*}{Selection} &\\
  \raisebox{0ex}{Decay mode}& scale & loss & \multirow{2}{*}{model} & \multirow{2}{*}{model} & \multirow{2}{*}{states} & &\raisebox{0ex}{Total}\\
  & calibration & correction &  &  &  & &\\
  \hline
 $J\mskip -3mu/\mskip -2mu\psi\mskip 2mu \pi^+$                   & 0.91 & 0.02 &0.10 &$\phantom{-}0.01$ &$<$0.01  &0.01 &0.92\\
 $J\mskip -3mu/\mskip -2mu\psi\mskip 2mu \pi^+ \pi^- \pi^+$             & 0.83 & 0.04 &0.10 &$\phantom{-}0.02$ &$<$0.01  &0.05 &0.84\\
 $J\mskip -3mu/\mskip -2mu\psi\mskip 2mu p \bar{p} \pi^+$          & 0.35 & 0.04 &0.10 &$\phantom{-}0.01$ &$<$0.01  &0.06 &0.37\\
 $J\mskip -3mu/\mskip -2mu\psi\mskip 2mu D_{s}^{+} (K^+ K^- \pi^+)$     & 0.36 & 0.04 &0.10 &$\phantom{-}0.02$ &$\phantom{-}0.07$   &0.02 &0.38\\
 $J\mskip -3mu/\mskip -2mu\psi\mskip 2mu D_{s}^{+} (\pi^+ \pi^- \pi^+)$ & 0.36 & 0.04 &0.10 &$\phantom{-}0.02$ &$\phantom{-}0.07$   &0.03 &0.38\\
 $J\mskip -3mu/\mskip -2mu\psi\mskip 2mu D^{0} (K^- \pi^+) K^{+}$       & 0.25 & 0.04 &0.10 &$\phantom{-}0.01$ &$\phantom{-}0.05$   &0.02 &0.28\\
 $B_{s}^{0} (D_s^- \pi^+) \pi^+$         & 0.23 & 0.04 &0.10 &$<$0.01    &$\phantom{-}0.21$   &0.12 &0.43\\
 $B_{s}^{0} (J\mskip -3mu/\mskip -2mu\psi\mskip 2mu \phi) \pi^{+}$          & 0.23 & 0.04 &0.10 &$\phantom{-}0.01$ &$\phantom{-}0.21$   &0.02 &0.41\\
\hline
\end{tabular}}
\end{center}
\end{table}

\begin{table}
\begin{center}
  \caption{\small \label{tab:Bs_systematics} Summary of systematic uncertainties on the mass difference $\Delta M$ (in $\mathrm{\,Me\kern -0.1em V}/c^{2}$) for the $B_{s}^{0} (D_s^- \pi^+) \pi^+$ and $B_{s}^{0} (J\mskip -3mu/\mskip -2mu\psi\mskip 2mu \phi) \pi^{+}$ decays.}
\resizebox{\textwidth}{!}{
\begin{tabular}{@{}l|cccccc|c@{}}
\hline
  & Momentum & \multirow{2}{*}{Energy} & \multirow{2}{*}{Signal} & \multirow{2}{*}{Background} & \multirow{2}{*}{Intermediate} & \multirow{3}{*}{Selection} & \\
 \raisebox{0ex}{Decay mode}& scale & \multirow{2}{*}{loss} & \multirow{2}{*}{model} & \multirow{2}{*}{model} & \multirow{2}{*}{states} & & \raisebox{0ex}{Total}\\
  & calibration &  &  &  &  & & \\
  \hline
 $B_{s}^{0} (D_s^- \pi^+) \pi^+$         & 0.23 & 0.04 &   0.10    &$\phantom{-}0.01$  &$<$0.01   &0.13 &0.29\\
 $B_{s}^{0} (J\mskip -3mu/\mskip -2mu\psi\mskip 2mu \phi) \pi^{+}$          & 0.23 & 0.04 &   0.10    &$<$0.01     &$<$0.01   &0.02 &0.25\\
\hline
\end{tabular}}
\end{center}
\end{table}

\section{Combination of the measurements}
\label{sec:combination}

The combination of the $B_{c}^{+}$ mass measurements
is performed using the Best Linear Unbiased Estimate (BLUE) method~\cite{Lyons:1988rp,Valassi:2003mu, Nisius:2020jmf}.
In the combination, uncertainties arising from the momentum-scale
calibration, energy loss corrections, and signal model are
assumed to be $100\%$ correlated, while all other sources of systematic
uncertainty are assumed to be uncorrelated. 
The uncertainty on the momentum-scale calibration of the $B_{s}^{0}$
mass ($0.14 \mathrm{\,Me\kern -0.1em V}/c^{2}$) is assumed to be $100\%$ correlated with that of the $B_{c}^{+}$ mass. 

\begin{figure}
  \centering
  \resizebox{0.8\textwidth}{!}{
    \includegraphics{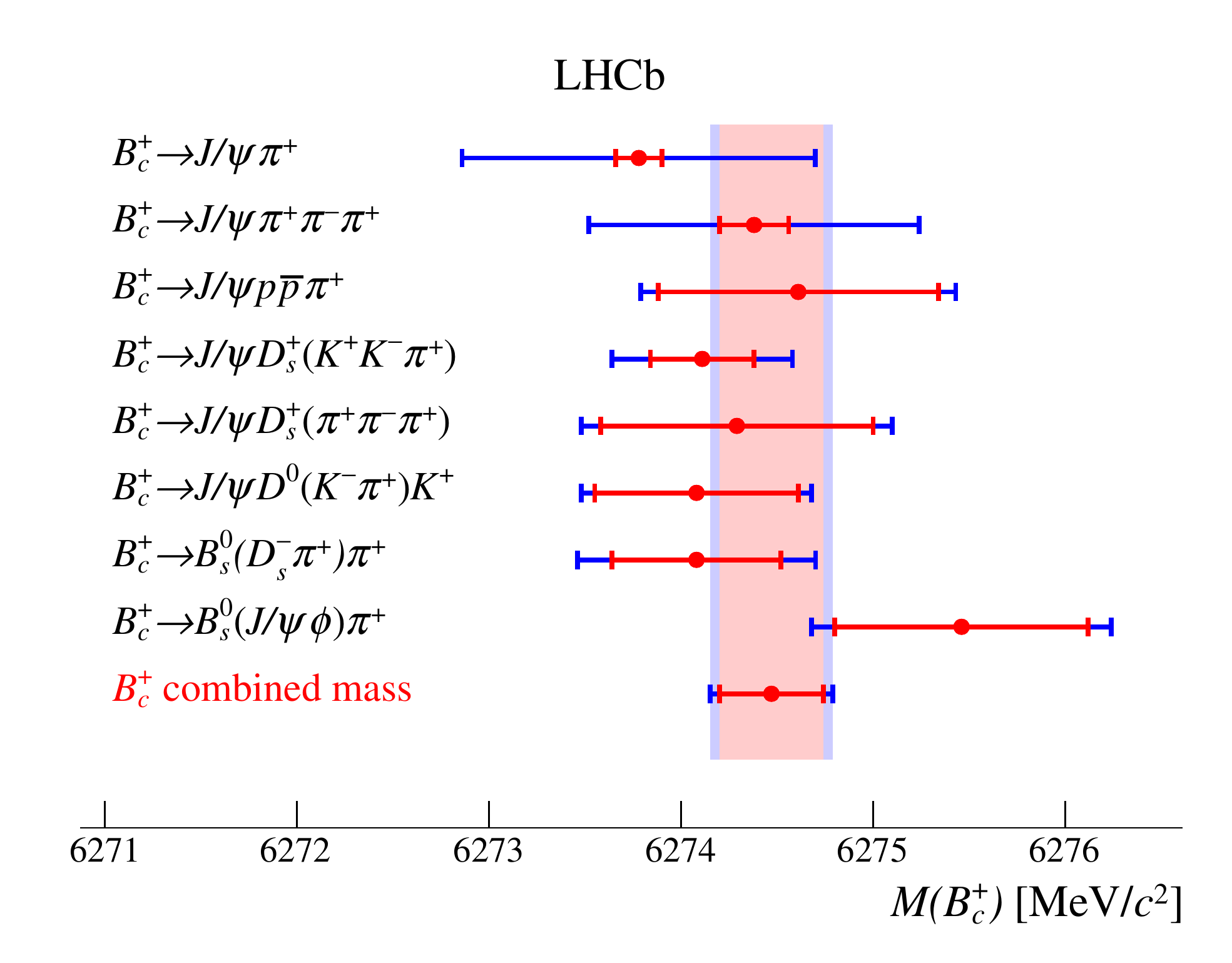}}
   \caption{Individual $B_{c}^{+}$ mass measurements and their combination. The red (inner) cross-bars show the statistical uncertainties, and the blue (outer) cross-bars show the total uncertainties.}
  \label{fig:combination}
\end{figure}

The individual mass measurements and the resulting combination are shown in Fig.~\ref{fig:combination}. 
The individual measurements are consistent with each other. 
The breakdown of the combined systematic uncertainty is
given in Table~\ref{tab:sum_uncertainty}.
The weights of individual measurements returned by the BLUE method are listed in
Table~\ref{tab:weights}.
The weights are computed including all uncertainties.
The measurement contributing most to the combination is obtained from the $B_{c}^{+} \rightarrow J\mskip -3mu/\mskip -2mu\psi\mskip 2mu D_{s}^{+} (K^+ K^- \pi^+)$ decay.
The negative weight for the $B_{c}^{+} \rightarrow J\mskip -3mu/\mskip -2mu\psi\mskip 2mu \pi^+$ channel arises from the  $100\%$ correlation between the systematic uncertainties due to the momentum-scale calibration. This results in a larger statistical and smaller systematic uncertainty relative to an uncorrelated average. 

\begin{figure}
  \centering
  \resizebox{0.8\textwidth}{!}{%
    \includegraphics{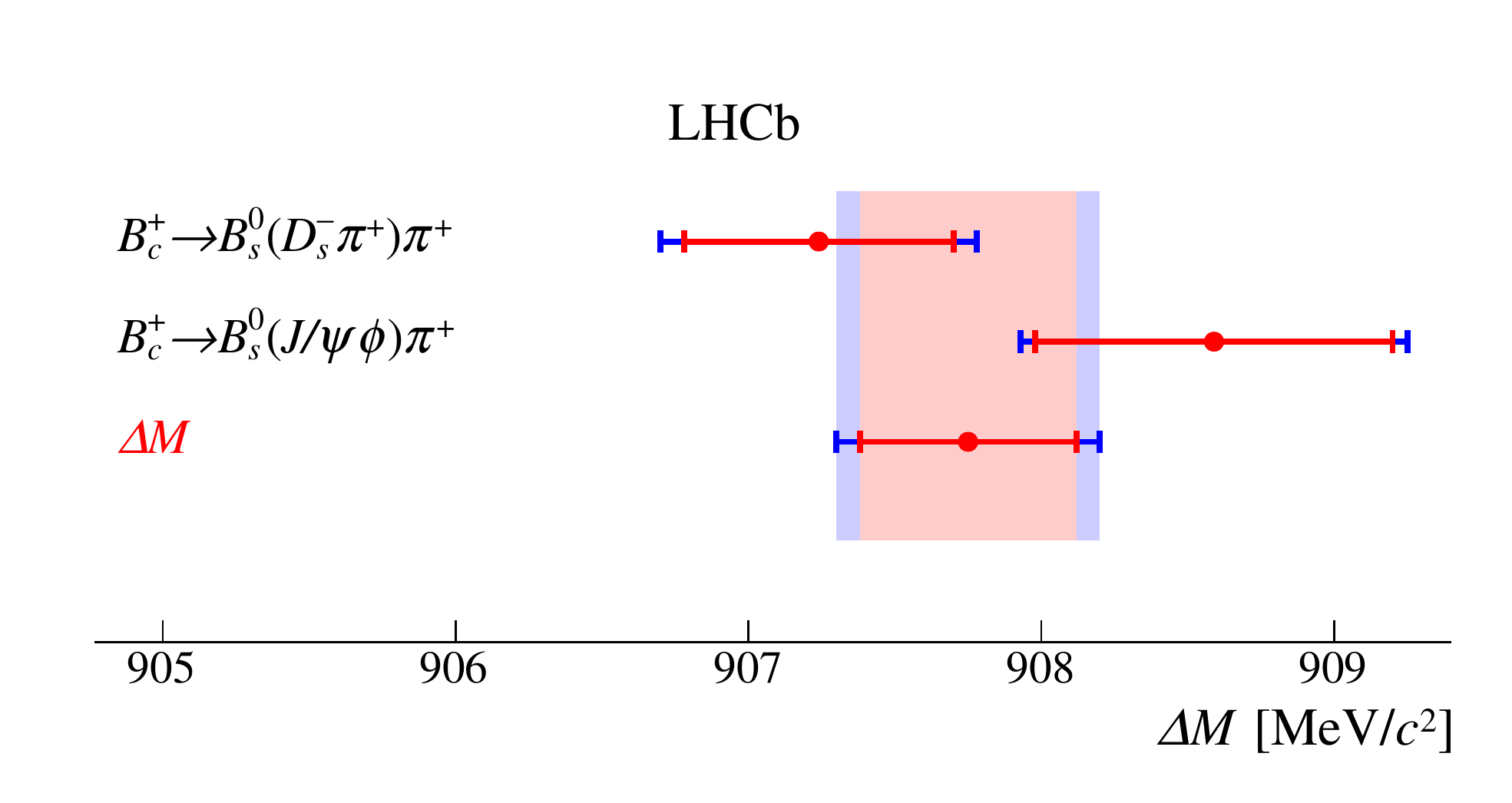}}
   \caption{Individual mass difference measurements and their combination. The red (inner) cross-bars show the statistical uncertainties, and the blue (outer) cross-bars show the total uncertainties on the measurement.}
  \label{fig:Bc_Bscombine}
\end{figure}

The combination for the mass difference $\Delta M$ is shown in Fig.~\ref{fig:Bc_Bscombine}.
The breakdown of the combined systematic uncertainty is
given in Table~\ref{tab:sum_uncertainty}
and the weights of decay modes in the combination are listed in
Table~\ref{tab:weights}.
The combined $B_{c}^{+}$ mass is determined to be \mbox{$M (B_{c}^{+}) = 6274.47 \pm 0.27 \,({\rm stat}) \pm 0.17 \,({\rm syst}) \mathrm{\,Me\kern -0.1em V}/c^{2}$},
while the mass difference between the $B_{c}^{+}$ and $B_{s}^{0}$ mesons, $\Delta M$, is determined to be \mbox{$\Delta M = 907.75 \pm 0.37 \,({\rm stat}) \pm 0.27 \,({\rm syst}) \mathrm{\,Me\kern -0.1em V}/c^{2}$}.

\begin{table}
\begin{center}
  \caption{\small \label{tab:sum_uncertainty} Breakdown of systematic uncertainties
    (in $\mathrm{\,Me\kern -0.1em V}/c^{2}$) in the combination of the $B_{c}^{+}$ mass and the mass difference $\Delta M$. The total uncertainty is the sum in quadrature of the uncertainty of different sources.}
\begin{tabular}{@{}l|cccccc@{}}
\hline
 Source & Mass & Mass difference \\
  \hline
 Momentum-scale calibration      & 0.11 &$\phantom{-}0.23$\\
 Energy loss                     & 0.05 &$\phantom{-}0.04$\\
 Signal line shape             & 0.10 &$\phantom{-}0.10$\\ 
 Background line shape          & 0.01 &$\phantom{-}0.01$\\
 Mass of intermediate state     & 0.06 &$<$0.01\\
 Selection bias correction     & 0.03 &$\phantom{-}0.08$\\
\hline
Total                         & 0.17 &$\phantom{-}0.27$\\
\hline
\end{tabular}
\end{center}
\end{table}

\begin{table}
\begin{center}
  \caption{\small \label{tab:weights} Weights of the decay modes in the
    combination of the $B_{c}^{+}$ mass and the mass difference $\Delta M$.}
\begin{tabular}{l|cc}
\hline
 Decay mode & Mass & Mass difference\\
  \hline
 $J\mskip -3mu/\mskip -2mu\psi\mskip 2mu \pi^+$                   & $-0.446$ & -\\
 $J\mskip -3mu/\mskip -2mu\psi\mskip 2mu \pi^+ \pi^- \pi^+$         & $\phantom{-}0.032$ & -\\
 $J\mskip -3mu/\mskip -2mu\psi\mskip 2mu p \bar{p} \pi^+$          & $\phantom{-}0.098$ & -\\
 $J\mskip -3mu/\mskip -2mu\psi\mskip 2mu D_{s}^{+} (K^+ K^- \pi^+)$     & $\phantom{-}0.659$ & -\\
 $J\mskip -3mu/\mskip -2mu\psi\mskip 2mu D_{s}^{+} (\pi^+ \pi^- \pi^+)$ & $\phantom{-}0.101$ & -\\
 $J\mskip -3mu/\mskip -2mu\psi\mskip 2mu D^{0} (K^- \pi^+) K^{+}$       & $\phantom{-}0.224$ & -\\
 $B_{s}^{0} (D_s^- \pi^+) \pi^+$         & $\phantom{-}0.220$ & 0.620\\
 $B_{s}^{0} (J\mskip -3mu/\mskip -2mu\psi\mskip 2mu \phi) \pi^{+}$          & $\phantom{-}0.111$ & 0.380\\
\hline
\end{tabular}
\end{center}
\end{table}

\section{Summary}
\label{sec:summary}

In summary, a precise measurement of the $B_{c}^{+}$ mass is performed 
using data samples collected in $pp$ collisions with the LHCb experiment at centre-of-mass energies 
of $\sqrt{s}$ = 7, 8 and 13 TeV, corresponding to an integrated
luminosity of $9$ fb$^{-1}$. 
The $B_{c}^{+}$ candidates are reconstructed via the decays
$B_{c}^{+} \rightarrow J\mskip -3mu/\mskip -2mu\psi\mskip 2mu \pi^+$,
$B_{c}^{+} \rightarrow J\mskip -3mu/\mskip -2mu\psi\mskip 2mu \pi^+ \pi^- \pi^+$,
$B_{c}^{+} \rightarrow J\mskip -3mu/\mskip -2mu\psi\mskip 2mu p \bar{p} \pi^+$,
$B_{c}^{+} \rightarrow J\mskip -3mu/\mskip -2mu\psi\mskip 2mu D_{s}^{+} (K^+ K^- \pi^+)$, 
$B_{c}^{+} \rightarrow J\mskip -3mu/\mskip -2mu\psi\mskip 2mu D_{s}^{+} (\pi^+ \pi^- \pi^+)$, 
$B_{c}^{+} \rightarrow J\mskip -3mu/\mskip -2mu\psi\mskip 2mu D^{0} (K^{-} \pi^{+}) K^{+}$,
$B_{c}^{+} \rightarrow B_{s}^{0} (D_s^- \pi^+) \pi^+$
and $B_{c}^{+} \rightarrow B_{s}^{0} (J\mskip -3mu/\mskip -2mu\psi\mskip 2mu \phi) \pi^+$. 
The $B_{c}^{+}$ mass is determined to be
\begin{equation}\label{mass_summary}
  6274.47 \pm 0.27 \,({\rm stat}) \pm 0.17 \,({\rm syst}) \mathrm{\,Me\kern -0.1em V}/c^{2}. \nonumber
\end{equation}
This result is consistent with theoretical predictions from perturbative and lattice QCD.
The mass difference between the $B_{c}^{+}$ and $B_{s}^{0}$ mesons, $\Delta M$, is determined to be
\begin{equation*}\label{Bsmass_summary}
  907.75 \pm 0.37 \,({\rm stat}) \pm 0.27 \,({\rm syst}) \mathrm{\,Me\kern -0.1em V}/c^{2}.
\end{equation*}
These results are the most accurate measurements of the $B_{c}^{+}$ mass to date. The precision compared to the world average~\cite{PDG2018} is improved by a factor of 2.

\section*{Acknowledgements}
%
%
\noindent We express our gratitude to our colleagues in the CERN
accelerator departments for the excellent performance of the LHC. We
thank the technical and administrative staff at the LHCb
institutes.
We acknowledge support from CERN and from the national agencies:
CAPES, CNPq, FAPERJ and FINEP (Brazil); 
MOST and NSFC (China); 
CNRS/IN2P3 (France); 
BMBF, DFG and MPG (Germany); 
INFN (Italy); 
NWO (Netherlands); 
MNiSW and NCN (Poland); 
MEN/IFA (Romania); 
MSHE (Russia); 
MinECo (Spain); 
SNSF and SER (Switzerland); 
NASU (Ukraine); 
STFC (United Kingdom); 
DOE NP and NSF (USA).
We acknowledge the computing resources that are provided by CERN, IN2P3
(France), KIT and DESY (Germany), INFN (Italy), SURF (Netherlands),
PIC (Spain), GridPP (United Kingdom), RRCKI and Yandex
LLC (Russia), CSCS (Switzerland), IFIN-HH (Romania), CBPF (Brazil),
PL-GRID (Poland) and OSC (USA).
We are indebted to the communities behind the multiple open-source
software packages on which we depend.
Individual groups or members have received support from
AvH Foundation (Germany);
EPLANET, Marie Sk\l{}odowska-Curie Actions and ERC (European Union);
ANR, Labex P2IO and OCEVU, and R\'{e}gion Auvergne-Rh\^{o}ne-Alpes (France);
Key Research Program of Frontier Sciences of CAS, CAS PIFI, and the Thousand Talents Program (China);
RFBR, RSF and Yandex LLC (Russia);
GVA, XuntaGal and GENCAT (Spain);
the Royal Society
and the Leverhulme Trust (United Kingdom).

\clearpage
\addcontentsline{toc}{section}{References}
\setboolean{inbibliography}{true}
\bibliographystyle{LHCb}
\bibliography{main,standard,BcReview,LHCb-PAPER,LHCb-CONF,LHCb-DP,LHCb-TDR}

\ifx\mcitethebibliography\mciteundefinedmacro
\PackageError{LHCb.bst}{mciteplus.sty has not been loaded}
{This bibstyle requires the use of the mciteplus package.}\fi
\providecommand{\href}[2]{#2}
\begin{mcitethebibliography}{10}
\mciteSetBstSublistMode{n}
\mciteSetBstMaxWidthForm{subitem}{\alph{mcitesubitemcount})}
\mciteSetBstSublistLabelBeginEnd{\mcitemaxwidthsubitemform\space}
{\relax}{\relax}

\bibitem{Gershtein:1987jj}
S.~S. Gershtein {\em et~al.},
  \ifthenelse{\boolean{articletitles}}{\emph{{Production cross-section and
  spectroscopy of $B_c$ mesons}}, }{}Sov.\ J.\ Nucl.\ Phys.\  \textbf{48}
  (1988) 327\relax
\mciteBstWouldAddEndPuncttrue
\mciteSetBstMidEndSepPunct{\mcitedefaultmidpunct}
{\mcitedefaultendpunct}{\mcitedefaultseppunct}\relax
\EndOfBibitem
\bibitem{Chen:1992fq}
Y.-Q. Chen and Y.-P. Kuang,
  \ifthenelse{\boolean{articletitles}}{\emph{{Improved QCD motivated heavy
  quark potentials with explicit $\Lambda_{\overline{\mathrm{MS}}}$
  dependence}}, }{}\href{https://doi.org/10.1103/PhysRevD.46.1165}{Phys.\ Rev.\
   \textbf{D46} (1992) 1165}\relax
\mciteBstWouldAddEndPuncttrue
\mciteSetBstMidEndSepPunct{\mcitedefaultmidpunct}
{\mcitedefaultendpunct}{\mcitedefaultseppunct}\relax
\EndOfBibitem
\bibitem{Eichten:1994gt}
E.~J. Eichten and C.~Quigg, \ifthenelse{\boolean{articletitles}}{\emph{{Mesons
  with beauty and charm: spectroscopy}},
  }{}\href{https://doi.org/10.1103/PhysRevD.49.5845}{Phys.\ Rev.\  \textbf{D49}
  (1994) 5845},
  \href{http://arxiv.org/abs/hep-ph/9402210}{{\normalfont\ttfamily
  arXiv:hep-ph/9402210}}\relax
\mciteBstWouldAddEndPuncttrue
\mciteSetBstMidEndSepPunct{\mcitedefaultmidpunct}
{\mcitedefaultendpunct}{\mcitedefaultseppunct}\relax
\EndOfBibitem
\bibitem{Kiselev:1994rc}
V.~V. Kiselev, A.~K. Likhoded, and A.~V. Tkabladze,
  \ifthenelse{\boolean{articletitles}}{\emph{{$B_c$ spectroscopy}},
  }{}\href{https://doi.org/10.1103/PhysRevD.51.3613}{Phys.\ Rev.\  \textbf{D51}
  (1995) 3613},
  \href{http://arxiv.org/abs/hep-ph/9406339}{{\normalfont\ttfamily
  arXiv:hep-ph/9406339}}\relax
\mciteBstWouldAddEndPuncttrue
\mciteSetBstMidEndSepPunct{\mcitedefaultmidpunct}
{\mcitedefaultendpunct}{\mcitedefaultseppunct}\relax
\EndOfBibitem
\bibitem{Gupta:1995ps}
S.~N. Gupta and J.~M. Johnson,
  \ifthenelse{\boolean{articletitles}}{\emph{{$B_c$ spectroscopy in a
  quantum-chromodynamic potential model}},
  }{}\href{https://doi.org/10.1103/PhysRevD.53.312}{Phys.\ Rev.\  \textbf{D53}
  (1996) 312}, \href{http://arxiv.org/abs/hep-ph/9511267}{{\normalfont\ttfamily
  arXiv:hep-ph/9511267}}\relax
\mciteBstWouldAddEndPuncttrue
\mciteSetBstMidEndSepPunct{\mcitedefaultmidpunct}
{\mcitedefaultendpunct}{\mcitedefaultseppunct}\relax
\EndOfBibitem
\bibitem{Fulcher:1998ka}
L.~P. Fulcher, \ifthenelse{\boolean{articletitles}}{\emph{{Phenomenological
  predictions of the properties of the $B_c$ system}},
  }{}\href{https://doi.org/10.1103/PhysRevD.60.074006}{Phys.\ Rev.\
  \textbf{D60} (1999) 074006},
  \href{http://arxiv.org/abs/hep-ph/9806444}{{\normalfont\ttfamily
  arXiv:hep-ph/9806444}}\relax
\mciteBstWouldAddEndPuncttrue
\mciteSetBstMidEndSepPunct{\mcitedefaultmidpunct}
{\mcitedefaultendpunct}{\mcitedefaultseppunct}\relax
\EndOfBibitem
\bibitem{Devlani:2014nda}
N.~Devlani, V.~Kher, and A.~K. Rai,
  \ifthenelse{\boolean{articletitles}}{\emph{{Masses and electromagnetic
  transitions of the B$_{c}$ mesons}},
  }{}\href{https://doi.org/10.1140/epja/i2014-14154-2}{Eur.\ Phys.\ J.\
  \textbf{A50} (2014) 154}\relax
\mciteBstWouldAddEndPuncttrue
\mciteSetBstMidEndSepPunct{\mcitedefaultmidpunct}
{\mcitedefaultendpunct}{\mcitedefaultseppunct}\relax
\EndOfBibitem
\bibitem{Soni:2017wvy}
N.~R. Soni {\em et~al.},
  \ifthenelse{\boolean{articletitles}}{\emph{{$Q\bar{Q}\,(Q\in \{b, c\})$
  spectroscopy using the Cornell potential}},
  }{}\href{https://doi.org/10.1140/epjc/s10052-018-6068-6}{Eur.\ Phys.\ J.\
  \textbf{C78} (2018) 592},
  \href{http://arxiv.org/abs/1707.07144}{{\normalfont\ttfamily
  arXiv:1707.07144}}\relax
\mciteBstWouldAddEndPuncttrue
\mciteSetBstMidEndSepPunct{\mcitedefaultmidpunct}
{\mcitedefaultendpunct}{\mcitedefaultseppunct}\relax
\EndOfBibitem
\bibitem{Wei:2010zza}
K.-W. Wei and X.-H. Guo, \ifthenelse{\boolean{articletitles}}{\emph{{Mass
  spectra of doubly heavy mesons in Regge phenomenology}},
  }{}\href{https://doi.org/10.1103/PhysRevD.81.076005}{Phys.\ Rev.\
  \textbf{D81} (2010) 076005}\relax
\mciteBstWouldAddEndPuncttrue
\mciteSetBstMidEndSepPunct{\mcitedefaultmidpunct}
{\mcitedefaultendpunct}{\mcitedefaultseppunct}\relax
\EndOfBibitem
\bibitem{Chen:2020ecu}
M.~Chen, L.~Chang, and Y.-x. Liu,
  \ifthenelse{\boolean{articletitles}}{\emph{{$B_{c}$ meson spectrum via
  Dyson-Schwinger equation and Bethe-Salpeter equation approach}},
  }{}\href{https://doi.org/10.1103/PhysRevD.101.056002}{Phys.\ Rev.\ D
  \textbf{101} (2020) 056002},
  \href{http://arxiv.org/abs/2001.00161}{{\normalfont\ttfamily
  arXiv:2001.00161}}\relax
\mciteBstWouldAddEndPuncttrue
\mciteSetBstMidEndSepPunct{\mcitedefaultmidpunct}
{\mcitedefaultendpunct}{\mcitedefaultseppunct}\relax
\EndOfBibitem
\bibitem{Brambilla:2000db}
N.~Brambilla and A.~Vairo, \ifthenelse{\boolean{articletitles}}{\emph{{$B_c$
  mass up to order $\alpha_s^4$}},
  }{}\href{https://doi.org/10.1103/PhysRevD.62.094019}{Phys.\ Rev.\
  \textbf{D62} (2000) 094019},
  \href{http://arxiv.org/abs/hep-ph/0002075}{{\normalfont\ttfamily
  arXiv:hep-ph/0002075}}\relax
\mciteBstWouldAddEndPuncttrue
\mciteSetBstMidEndSepPunct{\mcitedefaultmidpunct}
{\mcitedefaultendpunct}{\mcitedefaultseppunct}\relax
\EndOfBibitem
\bibitem{Xiao:2013lia}
Z.-J. Xiao and X.~Liu, \ifthenelse{\boolean{articletitles}}{\emph{{The two-body
  hadronic decays of $B_c$ meson in the perturbative QCD approach: A short
  review}}, }{}\href{https://doi.org/10.1007/s11434-014-0418-z}{Chin.\ Sci.\
  Bull.\  \textbf{59} (2014) 3748},
  \href{http://arxiv.org/abs/1401.0151}{{\normalfont\ttfamily
  arXiv:1401.0151}}\relax
\mciteBstWouldAddEndPuncttrue
\mciteSetBstMidEndSepPunct{\mcitedefaultmidpunct}
{\mcitedefaultendpunct}{\mcitedefaultseppunct}\relax
\EndOfBibitem
\bibitem{Ebert:2002pp}
D.~Ebert, R.~N. Faustov, and V.~O. Galkin,
  \ifthenelse{\boolean{articletitles}}{\emph{{Properties of heavy quarkonia and
  $B_c$ mesons in the relativistic quark model}},
  }{}\href{https://doi.org/10.1103/PhysRevD.67.014027}{Phys.\ Rev.\
  \textbf{D67} (2003) 014027},
  \href{http://arxiv.org/abs/hep-ph/0210381}{{\normalfont\ttfamily
  arXiv:hep-ph/0210381}}\relax
\mciteBstWouldAddEndPuncttrue
\mciteSetBstMidEndSepPunct{\mcitedefaultmidpunct}
{\mcitedefaultendpunct}{\mcitedefaultseppunct}\relax
\EndOfBibitem
\bibitem{Godfrey:2004ya}
S.~Godfrey, \ifthenelse{\boolean{articletitles}}{\emph{{Spectroscopy of $B_c$
  mesons in the relativized quark model}},
  }{}\href{https://doi.org/10.1103/PhysRevD.70.054017}{Phys.\ Rev.\
  \textbf{D70} (2004) 054017},
  \href{http://arxiv.org/abs/hep-ph/0406228}{{\normalfont\ttfamily
  arXiv:hep-ph/0406228}}\relax
\mciteBstWouldAddEndPuncttrue
\mciteSetBstMidEndSepPunct{\mcitedefaultmidpunct}
{\mcitedefaultendpunct}{\mcitedefaultseppunct}\relax
\EndOfBibitem
\bibitem{Ebert:2011jc}
D.~Ebert, R.~N. Faustov, and V.~O. Galkin,
  \ifthenelse{\boolean{articletitles}}{\emph{{Spectroscopy and Regge
  trajectories of heavy quarkonia and $B_{c}$ mesons}},
  }{}\href{https://doi.org/10.1140/epjc/s10052-011-1825-9}{Eur.\ Phys.\ J.\ C
  \textbf{71} (2011) 1825},
  \href{http://arxiv.org/abs/1111.0454}{{\normalfont\ttfamily
  arXiv:1111.0454}}\relax
\mciteBstWouldAddEndPuncttrue
\mciteSetBstMidEndSepPunct{\mcitedefaultmidpunct}
{\mcitedefaultendpunct}{\mcitedefaultseppunct}\relax
\EndOfBibitem
\bibitem{Fischer:2014cfa}
C.~S. Fischer, S.~Kubrak, and R.~Williams,
  \ifthenelse{\boolean{articletitles}}{\emph{{Spectra of heavy mesons in the
  Bethe-Salpeter approach}},
  }{}\href{https://doi.org/10.1140/epja/i2015-15010-7}{Eur.\ Phys.\ J.\
  \textbf{A51} (2015) 10},
  \href{http://arxiv.org/abs/1409.5076}{{\normalfont\ttfamily
  arXiv:1409.5076}}\relax
\mciteBstWouldAddEndPuncttrue
\mciteSetBstMidEndSepPunct{\mcitedefaultmidpunct}
{\mcitedefaultendpunct}{\mcitedefaultseppunct}\relax
\EndOfBibitem
\bibitem{Monteiro:2016rzi}
A.~P. Monteiro, M.~Bhat, and K.~B. Vijaya~Kumar,
  \ifthenelse{\boolean{articletitles}}{\emph{{$c\bar{b}$ spectrum and decay
  properties with coupled channel effects}},
  }{}\href{https://doi.org/10.1103/PhysRevD.95.054016}{Phys.\ Rev.\
  \textbf{D95} (2017) 054016},
  \href{http://arxiv.org/abs/1608.05782}{{\normalfont\ttfamily
  arXiv:1608.05782}}\relax
\mciteBstWouldAddEndPuncttrue
\mciteSetBstMidEndSepPunct{\mcitedefaultmidpunct}
{\mcitedefaultendpunct}{\mcitedefaultseppunct}\relax
\EndOfBibitem
\bibitem{Davies:1996gi}
C.~T.~H. Davies {\em et~al.}, \ifthenelse{\boolean{articletitles}}{\emph{{$B_c$
  Spectroscopy from lattice QCD}},
  }{}\href{https://doi.org/10.1016/0370-2693(96)00650-8}{Phys.\ Lett.\
  \textbf{B382} (1996) 131},
  \href{http://arxiv.org/abs/hep-lat/9602020}{{\normalfont\ttfamily
  arXiv:hep-lat/9602020}}\relax
\mciteBstWouldAddEndPuncttrue
\mciteSetBstMidEndSepPunct{\mcitedefaultmidpunct}
{\mcitedefaultendpunct}{\mcitedefaultseppunct}\relax
\EndOfBibitem
\bibitem{Shanahan:1999mv}
UKQCD collaboration, H.~P. Shanahan, P.~Boyle, C.~T.~H. Davies, and H.~Newton,
  \ifthenelse{\boolean{articletitles}}{\emph{{A non-perturbative calculation of
  the mass of the $B_c$}},
  }{}\href{https://doi.org/10.1016/S0370-2693(99)00325-1}{Phys.\ Lett.\
  \textbf{B453} (1999) 289},
  \href{http://arxiv.org/abs/hep-lat/9902025}{{\normalfont\ttfamily
  arXiv:hep-lat/9902025}}\relax
\mciteBstWouldAddEndPuncttrue
\mciteSetBstMidEndSepPunct{\mcitedefaultmidpunct}
{\mcitedefaultendpunct}{\mcitedefaultseppunct}\relax
\EndOfBibitem
\bibitem{Allison:2004hy}
HPQCD collaboration, I.~F. Allison {\em et~al.},
  \ifthenelse{\boolean{articletitles}}{\emph{{A precise determination of the
  $B_c$ mass from dynamical lattice QCD}},
  }{}\href{https://doi.org/10.1016/j.nuclphysbps.2004.11.375}{Nucl.\ Phys.\
  Proc.\ Suppl.\  \textbf{140} (2005) 440},
  \href{http://arxiv.org/abs/hep-lat/0409090}{{\normalfont\ttfamily
  arXiv:hep-lat/0409090}}\relax
\mciteBstWouldAddEndPuncttrue
\mciteSetBstMidEndSepPunct{\mcitedefaultmidpunct}
{\mcitedefaultendpunct}{\mcitedefaultseppunct}\relax
\EndOfBibitem
\bibitem{Allison:2004be}
HPQCD collaboration, I.~F. Allison {\em et~al.},
  \ifthenelse{\boolean{articletitles}}{\emph{{Mass of the $B_c$ meson in
  three-flavor lattice QCD}},
  }{}\href{https://doi.org/10.1103/PhysRevLett.94.172001}{Phys.\ Rev.\ Lett.\
  \textbf{94} (2005) 172001},
  \href{http://arxiv.org/abs/hep-lat/0411027}{{\normalfont\ttfamily
  arXiv:hep-lat/0411027}}\relax
\mciteBstWouldAddEndPuncttrue
\mciteSetBstMidEndSepPunct{\mcitedefaultmidpunct}
{\mcitedefaultendpunct}{\mcitedefaultseppunct}\relax
\EndOfBibitem
\bibitem{Chiu:2007bc}
TWQCD collaboration, T.-W. Chiu and T.-H. Hsieh,
  \ifthenelse{\boolean{articletitles}}{\emph{{$B_{s}$ and $B_{c}$ mesons in
  lattice QCD with exact chiral symmetry}}, }{}PoS \textbf{LAT2006} (2007) 180,
  \href{http://arxiv.org/abs/0704.3495}{{\normalfont\ttfamily
  arXiv:0704.3495}}\relax
\mciteBstWouldAddEndPuncttrue
\mciteSetBstMidEndSepPunct{\mcitedefaultmidpunct}
{\mcitedefaultendpunct}{\mcitedefaultseppunct}\relax
\EndOfBibitem
\bibitem{Dowdall:2012ab}
R.~J. Dowdall, C.~T.~H. Davies, T.~C. Hammant, and R.~R. Horgan,
  \ifthenelse{\boolean{articletitles}}{\emph{{Precise heavy-light meson masses
  and hyperfine splittings from lattice QCD including charm quarks in the
  sea}}, }{}\href{https://doi.org/10.1103/PhysRevD.86.094510}{Phys.\ Rev.\
  \textbf{D86} (2012) 094510},
  \href{http://arxiv.org/abs/1207.5149}{{\normalfont\ttfamily
  arXiv:1207.5149}}\relax
\mciteBstWouldAddEndPuncttrue
\mciteSetBstMidEndSepPunct{\mcitedefaultmidpunct}
{\mcitedefaultendpunct}{\mcitedefaultseppunct}\relax
\EndOfBibitem
\bibitem{Abe:1998wi}
CDF collaboration, F.~Abe {\em et~al.},
  \ifthenelse{\boolean{articletitles}}{\emph{{Observation of the $B_c$ meson in
  $p\bar{p}$ collisions at \mbox{$\sqrt{s} = 1.8$ {\rm TeV}}}},
  }{}\href{https://doi.org/10.1103/PhysRevLett.81.2432}{Phys.\ Rev.\ Lett.\
  \textbf{81} (1998) 2432},
  \href{http://arxiv.org/abs/hep-ex/9805034}{{\normalfont\ttfamily
  arXiv:hep-ex/9805034}}\relax
\mciteBstWouldAddEndPuncttrue
\mciteSetBstMidEndSepPunct{\mcitedefaultmidpunct}
{\mcitedefaultendpunct}{\mcitedefaultseppunct}\relax
\EndOfBibitem
\bibitem{Abe:1998fb}
CDF collaboration, F.~Abe {\em et~al.},
  \ifthenelse{\boolean{articletitles}}{\emph{{Observation of $B_c$ mesons in
  $p\bar{p}$ collisions at \mbox{$\sqrt{s} = 1.8$ {\rm TeV}}}},
  }{}\href{https://doi.org/10.1103/PhysRevD.58.112004}{Phys.\ Rev.\
  \textbf{D58} (1998) 112004},
  \href{http://arxiv.org/abs/hep-ex/9804014}{{\normalfont\ttfamily
  arXiv:hep-ex/9804014}}\relax
\mciteBstWouldAddEndPuncttrue
\mciteSetBstMidEndSepPunct{\mcitedefaultmidpunct}
{\mcitedefaultendpunct}{\mcitedefaultseppunct}\relax
\EndOfBibitem
\bibitem{LHCb-PAPER-2012-028}
LHCb collaboration, R.~Aaij {\em et~al.},
  \ifthenelse{\boolean{articletitles}}{\emph{{Measurements of \Bcp production
  and mass with the \mbox{\decay{\Bc}{\jpsi\pip}} decay}},
  }{}\href{https://doi.org/10.1103/PhysRevLett.109.232001}{Phys.\ Rev.\ Lett.\
  \textbf{109} (2012) 232001},
  \href{http://arxiv.org/abs/1209.5634}{{\normalfont\ttfamily
  arXiv:1209.5634}}\relax
\mciteBstWouldAddEndPuncttrue
\mciteSetBstMidEndSepPunct{\mcitedefaultmidpunct}
{\mcitedefaultendpunct}{\mcitedefaultseppunct}\relax
\EndOfBibitem
\bibitem{LHCb-PAPER-2013-044}
LHCb collaboration, R.~Aaij {\em et~al.},
  \ifthenelse{\boolean{articletitles}}{\emph{{Observation of the decay
  \mbox{\decay{\Bc}{\Bs\pip}}}},
  }{}\href{https://doi.org/10.1103/PhysRevLett.111.181801}{Phys.\ Rev.\ Lett.\
  \textbf{111} (2013) 181801},
  \href{http://arxiv.org/abs/1308.4544}{{\normalfont\ttfamily
  arXiv:1308.4544}}\relax
\mciteBstWouldAddEndPuncttrue
\mciteSetBstMidEndSepPunct{\mcitedefaultmidpunct}
{\mcitedefaultendpunct}{\mcitedefaultseppunct}\relax
\EndOfBibitem
\bibitem{LHCb-PAPER-2019-033}
LHCb collaboration, R.~Aaij {\em et~al.},
  \ifthenelse{\boolean{articletitles}}{\emph{{Measurement of the \Bcm
  production fraction and asymmetry in 7 and 13 {\rm TeV} $pp$ collisions}},
  }{}\href{https://doi.org/10.1103/PhysRevD.100.112006}{Phys.\ Rev.\
  \textbf{D100} (2019) 112006},
  \href{http://arxiv.org/abs/1910.13404}{{\normalfont\ttfamily
  arXiv:1910.13404}}\relax
\mciteBstWouldAddEndPuncttrue
\mciteSetBstMidEndSepPunct{\mcitedefaultmidpunct}
{\mcitedefaultendpunct}{\mcitedefaultseppunct}\relax
\EndOfBibitem
\bibitem{LHCb-PAPER-2014-050}
LHCb collaboration, R.~Aaij {\em et~al.},
  \ifthenelse{\boolean{articletitles}}{\emph{{Measurement of \Bcp production in
  proton-proton collisions at \mbox{$\sqs=$8 {\rm TeV}}}},
  }{}\href{https://doi.org/10.1103/PhysRevLett.114.132001}{Phys.\ Rev.\ Lett.\
  \textbf{114} (2015) 132001},
  \href{http://arxiv.org/abs/1411.2943}{{\normalfont\ttfamily
  arXiv:1411.2943}}\relax
\mciteBstWouldAddEndPuncttrue
\mciteSetBstMidEndSepPunct{\mcitedefaultmidpunct}
{\mcitedefaultendpunct}{\mcitedefaultseppunct}\relax
\EndOfBibitem
\bibitem{LHCb-PAPER-2016-058}
LHCb collaboration, R.~Aaij {\em et~al.},
  \ifthenelse{\boolean{articletitles}}{\emph{{Observation of
  \mbox{\decay{\Bc}{\Dz\Kp}} decays}},
  }{}\href{https://doi.org/10.1103/PhysRevLett.118.111803}{Phys.\ Rev.\ Lett.\
  \textbf{118} (2017) 111803},
  \href{http://arxiv.org/abs/1701.01856}{{\normalfont\ttfamily
  arXiv:1701.01856}}\relax
\mciteBstWouldAddEndPuncttrue
\mciteSetBstMidEndSepPunct{\mcitedefaultmidpunct}
{\mcitedefaultendpunct}{\mcitedefaultseppunct}\relax
\EndOfBibitem
\bibitem{LHCb-PAPER-2013-010}
LHCb collaboration, R.~Aaij {\em et~al.},
  \ifthenelse{\boolean{articletitles}}{\emph{{Observation of
  \mbox{\decay{\Bc}{\jpsi\Dsp}} and \mbox{\decay{\Bc}{\jpsi\Dssp}} decays}},
  }{}\href{https://doi.org/10.1103/PhysRevD.87.112012}{Phys.\ Rev.\
  \textbf{D87} (2013) 112012},
  \href{http://arxiv.org/abs/1304.4530}{{\normalfont\ttfamily
  arXiv:1304.4530}}\relax
\mciteBstWouldAddEndPuncttrue
\mciteSetBstMidEndSepPunct{\mcitedefaultmidpunct}
{\mcitedefaultendpunct}{\mcitedefaultseppunct}\relax
\EndOfBibitem
\bibitem{LHCb-PAPER-2014-039}
LHCb collaboration, R.~Aaij {\em et~al.},
  \ifthenelse{\boolean{articletitles}}{\emph{{First observation of a baryonic
  \Bcp decay}}, }{}\href{https://doi.org/10.1103/PhysRevLett.113.152003}{Phys.\
  Rev.\ Lett.\  \textbf{113} (2014) 152003},
  \href{http://arxiv.org/abs/1408.0971}{{\normalfont\ttfamily
  arXiv:1408.0971}}\relax
\mciteBstWouldAddEndPuncttrue
\mciteSetBstMidEndSepPunct{\mcitedefaultmidpunct}
{\mcitedefaultendpunct}{\mcitedefaultseppunct}\relax
\EndOfBibitem
\bibitem{LHCb-PAPER-2016-055}
LHCb collaboration, R.~Aaij {\em et~al.},
  \ifthenelse{\boolean{articletitles}}{\emph{{Observation of
  \mbox{\decay{\Bc}{\jpsi D^{(\ast)} K^{(\ast)}}} decays}},
  }{}\href{https://doi.org/10.1103/PhysRevD.95.032005}{Phys.\ Rev.\
  \textbf{D95} (2017) 032005},
  \href{http://arxiv.org/abs/1612.07421}{{\normalfont\ttfamily
  arXiv:1612.07421}}\relax
\mciteBstWouldAddEndPuncttrue
\mciteSetBstMidEndSepPunct{\mcitedefaultmidpunct}
{\mcitedefaultendpunct}{\mcitedefaultseppunct}\relax
\EndOfBibitem
\bibitem{LHCb-PAPER-2019-007}
LHCb collaboration, R.~Aaij {\em et~al.},
  \ifthenelse{\boolean{articletitles}}{\emph{{Observation of an excited \Bc
  state}}, }{}\href{https://doi.org/10.1103/PhysRevLett.122.232001}{Phys.\
  Rev.\ Lett.\  \textbf{122} (2019) 232001},
  \href{http://arxiv.org/abs/1904.00081}{{\normalfont\ttfamily
  arXiv:1904.00081}}\relax
\mciteBstWouldAddEndPuncttrue
\mciteSetBstMidEndSepPunct{\mcitedefaultmidpunct}
{\mcitedefaultendpunct}{\mcitedefaultseppunct}\relax
\EndOfBibitem
\bibitem{LHCb-PAPER-2013-063}
LHCb collaboration, R.~Aaij {\em et~al.},
  \ifthenelse{\boolean{articletitles}}{\emph{{Measurement of the \Bcp meson
  lifetime using \mbox{\decay{\Bc}{\jpsi\mup\neum X}} decays}},
  }{}\href{https://doi.org/10.1140/epjc/s10052-014-2839-x}{Eur.\ Phys.\ J.\
  \textbf{C74} (2014) 2839},
  \href{http://arxiv.org/abs/1401.6932}{{\normalfont\ttfamily
  arXiv:1401.6932}}\relax
\mciteBstWouldAddEndPuncttrue
\mciteSetBstMidEndSepPunct{\mcitedefaultmidpunct}
{\mcitedefaultendpunct}{\mcitedefaultseppunct}\relax
\EndOfBibitem
\bibitem{LHCb-PAPER-2014-060}
LHCb collaboration, R.~Aaij {\em et~al.},
  \ifthenelse{\boolean{articletitles}}{\emph{{Measurement of the lifetime of
  the \Bcp meson using the \mbox{\decay{\Bc}{\jpsi\pip}} decay mode}},
  }{}\href{https://doi.org/10.1016/j.physletb.2015.01.010}{Phys.\ Lett.\
  \textbf{B742} (2015) 29},
  \href{http://arxiv.org/abs/1411.6899}{{\normalfont\ttfamily
  arXiv:1411.6899}}\relax
\mciteBstWouldAddEndPuncttrue
\mciteSetBstMidEndSepPunct{\mcitedefaultmidpunct}
{\mcitedefaultendpunct}{\mcitedefaultseppunct}\relax
\EndOfBibitem
\bibitem{LHCb-PAPER-2012-054}
LHCb collaboration, R.~Aaij {\em et~al.},
  \ifthenelse{\boolean{articletitles}}{\emph{{Observation of the decay
  \mbox{\decay{\Bc}{\psitwos\pip}}}},
  }{}\href{https://doi.org/10.1103/PhysRevD.87.071103}{Phys.\ Rev.\
  \textbf{D87} (2013) 071103(R)},
  \href{http://arxiv.org/abs/1303.1737}{{\normalfont\ttfamily
  arXiv:1303.1737}}\relax
\mciteBstWouldAddEndPuncttrue
\mciteSetBstMidEndSepPunct{\mcitedefaultmidpunct}
{\mcitedefaultendpunct}{\mcitedefaultseppunct}\relax
\EndOfBibitem
\bibitem{LHCb-PAPER-2013-021}
LHCb collaboration, R.~Aaij {\em et~al.},
  \ifthenelse{\boolean{articletitles}}{\emph{{First observation of the decay
  \mbox{\decay{\Bc}{\jpsi\Kp}}}},
  }{}\href{https://doi.org/10.1007/JHEP09(2013)075}{JHEP \textbf{09} (2013)
  075}, \href{http://arxiv.org/abs/1306.6723}{{\normalfont\ttfamily
  arXiv:1306.6723}}\relax
\mciteBstWouldAddEndPuncttrue
\mciteSetBstMidEndSepPunct{\mcitedefaultmidpunct}
{\mcitedefaultendpunct}{\mcitedefaultseppunct}\relax
\EndOfBibitem
\bibitem{LHCb-PAPER-2013-047}
LHCb collaboration, R.~Aaij {\em et~al.},
  \ifthenelse{\boolean{articletitles}}{\emph{{Observation of the decay
  \mbox{\decay{\Bc}{\jpsi\Kp\Km\pip}}}},
  }{}\href{https://doi.org/10.1007/JHEP11(2013)094}{JHEP \textbf{11} (2013)
  094}, \href{http://arxiv.org/abs/1309.0587}{{\normalfont\ttfamily
  arXiv:1309.0587}}\relax
\mciteBstWouldAddEndPuncttrue
\mciteSetBstMidEndSepPunct{\mcitedefaultmidpunct}
{\mcitedefaultendpunct}{\mcitedefaultseppunct}\relax
\EndOfBibitem
\bibitem{LHCb-PAPER-2015-024}
LHCb collaboration, R.~Aaij {\em et~al.},
  \ifthenelse{\boolean{articletitles}}{\emph{{Measurement of the branching
  fraction ratio
  \mbox{$\BF(\decay{\Bcp}{\psitwos\pip})/\BF(\decay{\Bcp}{\jpsi\pip})$}}},
  }{}\href{https://doi.org/10.1103/PhysRevD.92.072007}{Phys.\ Rev.\
  \textbf{D92} (2015) 057007},
  \href{http://arxiv.org/abs/1507.03516}{{\normalfont\ttfamily
  arXiv:1507.03516}}\relax
\mciteBstWouldAddEndPuncttrue
\mciteSetBstMidEndSepPunct{\mcitedefaultmidpunct}
{\mcitedefaultendpunct}{\mcitedefaultseppunct}\relax
\EndOfBibitem
\bibitem{LHCb-PAPER-2016-020}
LHCb collaboration, R.~Aaij {\em et~al.},
  \ifthenelse{\boolean{articletitles}}{\emph{{Measurement of the ratio of
  branching fractions
  \mbox{$\BF(\decay{\Bcp}{\jpsi\Kp})/\BF(\decay{\Bcp}{\jpsi\pip})$}}},
  }{}\href{https://doi.org/10.1007/JHEP09(2016)153}{JHEP \textbf{09} (2016)
  153}, \href{http://arxiv.org/abs/1607.06823}{{\normalfont\ttfamily
  arXiv:1607.06823}}\relax
\mciteBstWouldAddEndPuncttrue
\mciteSetBstMidEndSepPunct{\mcitedefaultmidpunct}
{\mcitedefaultendpunct}{\mcitedefaultseppunct}\relax
\EndOfBibitem
\bibitem{LHCb-PAPER-2016-001}
LHCb collaboration, R.~Aaij {\em et~al.},
  \ifthenelse{\boolean{articletitles}}{\emph{{Search for \Bcp decays to the
  $\proton\antiproton\pip$ final state}},
  }{}\href{https://doi.org/10.1016/j.physletb.2016.05.074}{Phys.\ Lett.\
  \textbf{B759} (2016) 313},
  \href{http://arxiv.org/abs/1603.07037}{{\normalfont\ttfamily
  arXiv:1603.07037}}\relax
\mciteBstWouldAddEndPuncttrue
\mciteSetBstMidEndSepPunct{\mcitedefaultmidpunct}
{\mcitedefaultendpunct}{\mcitedefaultseppunct}\relax
\EndOfBibitem
\bibitem{LHCb-PAPER-2016-022}
LHCb collaboration, R.~Aaij {\em et~al.},
  \ifthenelse{\boolean{articletitles}}{\emph{{Study of \Bcp decays to the
  $\Kp\Km\pip$ final state and evidence for the decay
  \mbox{\decay{\Bc}{\chiczero\pip}}}},
  }{}\href{https://doi.org/10.1103/PhysRevD.94.091102}{Phys.\ Rev.\
  \textbf{D94} (2016) 091102(R)},
  \href{http://arxiv.org/abs/1607.06134}{{\normalfont\ttfamily
  arXiv:1607.06134}}\relax
\mciteBstWouldAddEndPuncttrue
\mciteSetBstMidEndSepPunct{\mcitedefaultmidpunct}
{\mcitedefaultendpunct}{\mcitedefaultseppunct}\relax
\EndOfBibitem
\bibitem{LHCb-PAPER-2017-035}
LHCb collaboration, R.~Aaij {\em et~al.},
  \ifthenelse{\boolean{articletitles}}{\emph{{Measurement of the ratio of
  branching fractions
  \mbox{$\mathcal{B}(\decay{\Bc}{\jpsi\taup\nu_{\tau}})/\mathcal{B}(\decay{\Bc}{\jpsi\mup\nu_{\mu}})$}}},
  }{}\href{https://doi.org/10.1103/PhysRevLett.120.121801}{Phys.\ Rev.\ Lett.\
  \textbf{120} (2018) 121801},
  \href{http://arxiv.org/abs/1711.05623}{{\normalfont\ttfamily
  arXiv:1711.05623}}\relax
\mciteBstWouldAddEndPuncttrue
\mciteSetBstMidEndSepPunct{\mcitedefaultmidpunct}
{\mcitedefaultendpunct}{\mcitedefaultseppunct}\relax
\EndOfBibitem
\bibitem{LHCb-PAPER-2017-045}
LHCb collaboration, R.~Aaij {\em et~al.},
  \ifthenelse{\boolean{articletitles}}{\emph{{Search for \Bc decays to two
  charm mesons}},
  }{}\href{https://doi.org/10.1016/j.nuclphysb.2018.03.015}{Nucl.\ Phys.\
  \textbf{B930} (2018) 563},
  \href{http://arxiv.org/abs/1712.04702}{{\normalfont\ttfamily
  arXiv:1712.04702}}\relax
\mciteBstWouldAddEndPuncttrue
\mciteSetBstMidEndSepPunct{\mcitedefaultmidpunct}
{\mcitedefaultendpunct}{\mcitedefaultseppunct}\relax
\EndOfBibitem
\bibitem{PDG2018}
Particle Data Group, M.~Tanabashi {\em et~al.},
  \ifthenelse{\boolean{articletitles}}{\emph{{\href{http://pdg.lbl.gov/}{Review
  of particle physics}}},
  }{}\href{https://doi.org/10.1103/PhysRevD.98.030001}{Phys.\ Rev.\
  \textbf{D98} (2018) 030001}\relax
\mciteBstWouldAddEndPuncttrue
\mciteSetBstMidEndSepPunct{\mcitedefaultmidpunct}
{\mcitedefaultendpunct}{\mcitedefaultseppunct}\relax
\EndOfBibitem
\bibitem{Sirunyan:2019osb}
CMS collaboration, A.~M. Sirunyan {\em et~al.},
  \ifthenelse{\boolean{articletitles}}{\emph{{Observation of two excited
  B$^+_\mathrm{c}$ states and measurement of the B$^+_\mathrm{c}$(2S) mass in
  pp collisions at $\sqrt{s} =$ 13 ${\rm TeV}$}},
  }{}\href{https://doi.org/10.1103/PhysRevLett.122.132001}{Phys.\ Rev.\ Lett.\
  \textbf{122} (2019) 132001},
  \href{http://arxiv.org/abs/1902.00571}{{\normalfont\ttfamily
  arXiv:1902.00571}}\relax
\mciteBstWouldAddEndPuncttrue
\mciteSetBstMidEndSepPunct{\mcitedefaultmidpunct}
{\mcitedefaultendpunct}{\mcitedefaultseppunct}\relax
\EndOfBibitem
\bibitem{LHCb-DP-2008-001}
LHCb collaboration, A.~A. Alves~Jr.\ {\em et~al.},
  \ifthenelse{\boolean{articletitles}}{\emph{{The \lhcb detector at the LHC}},
  }{}\href{https://doi.org/10.1088/1748-0221/3/08/S08005}{JINST \textbf{3}
  (2008) S08005}\relax
\mciteBstWouldAddEndPuncttrue
\mciteSetBstMidEndSepPunct{\mcitedefaultmidpunct}
{\mcitedefaultendpunct}{\mcitedefaultseppunct}\relax
\EndOfBibitem
\bibitem{LHCb-DP-2014-002}
LHCb collaboration, R.~Aaij {\em et~al.},
  \ifthenelse{\boolean{articletitles}}{\emph{{LHCb detector performance}},
  }{}\href{https://doi.org/10.1142/S0217751X15300227}{Int.\ J.\ Mod.\ Phys.\
  \textbf{A30} (2015) 1530022},
  \href{http://arxiv.org/abs/1412.6352}{{\normalfont\ttfamily
  arXiv:1412.6352}}\relax
\mciteBstWouldAddEndPuncttrue
\mciteSetBstMidEndSepPunct{\mcitedefaultmidpunct}
{\mcitedefaultendpunct}{\mcitedefaultseppunct}\relax
\EndOfBibitem
\bibitem{LHCb-DP-2014-001}
R.~Aaij {\em et~al.}, \ifthenelse{\boolean{articletitles}}{\emph{{Performance
  of the LHCb Vertex Locator}},
  }{}\href{https://doi.org/10.1088/1748-0221/9/09/P09007}{JINST \textbf{9}
  (2014) P09007}, \href{http://arxiv.org/abs/1405.7808}{{\normalfont\ttfamily
  arXiv:1405.7808}}\relax
\mciteBstWouldAddEndPuncttrue
\mciteSetBstMidEndSepPunct{\mcitedefaultmidpunct}
{\mcitedefaultendpunct}{\mcitedefaultseppunct}\relax
\EndOfBibitem
\bibitem{LHCb-DP-2013-003}
R.~Arink {\em et~al.}, \ifthenelse{\boolean{articletitles}}{\emph{{Performance
  of the LHCb Outer Tracker}},
  }{}\href{https://doi.org/10.1088/1748-0221/9/01/P01002}{JINST \textbf{9}
  (2014) P01002}, \href{http://arxiv.org/abs/1311.3893}{{\normalfont\ttfamily
  arXiv:1311.3893}}\relax
\mciteBstWouldAddEndPuncttrue
\mciteSetBstMidEndSepPunct{\mcitedefaultmidpunct}
{\mcitedefaultendpunct}{\mcitedefaultseppunct}\relax
\EndOfBibitem
\bibitem{LHCb-DP-2017-001}
P.~d'Argent {\em et~al.}, \ifthenelse{\boolean{articletitles}}{\emph{{Improved
  performance of the LHCb Outer Tracker in LHC Run 2}},
  }{}\href{https://doi.org/10.1088/1748-0221/12/11/P11016}{JINST \textbf{12}
  (2017) P11016}, \href{http://arxiv.org/abs/1708.00819}{{\normalfont\ttfamily
  arXiv:1708.00819}}\relax
\mciteBstWouldAddEndPuncttrue
\mciteSetBstMidEndSepPunct{\mcitedefaultmidpunct}
{\mcitedefaultendpunct}{\mcitedefaultseppunct}\relax
\EndOfBibitem
\bibitem{LHCb-PAPER-2012-048}
LHCb collaboration, R.~Aaij {\em et~al.},
  \ifthenelse{\boolean{articletitles}}{\emph{{Measurements of the \Lb, \Xibm,
  and \Omegab baryon masses}},
  }{}\href{https://doi.org/10.1103/PhysRevLett.110.182001}{Phys.\ Rev.\ Lett.\
  \textbf{110} (2013) 182001},
  \href{http://arxiv.org/abs/1302.1072}{{\normalfont\ttfamily
  arXiv:1302.1072}}\relax
\mciteBstWouldAddEndPuncttrue
\mciteSetBstMidEndSepPunct{\mcitedefaultmidpunct}
{\mcitedefaultendpunct}{\mcitedefaultseppunct}\relax
\EndOfBibitem
\bibitem{LHCb-PAPER-2013-011}
LHCb collaboration, R.~Aaij {\em et~al.},
  \ifthenelse{\boolean{articletitles}}{\emph{{Precision measurement of \D meson
  mass differences}}, }{}\href{https://doi.org/10.1007/JHEP06(2013)065}{JHEP
  \textbf{06} (2013) 065},
  \href{http://arxiv.org/abs/1304.6865}{{\normalfont\ttfamily
  arXiv:1304.6865}}\relax
\mciteBstWouldAddEndPuncttrue
\mciteSetBstMidEndSepPunct{\mcitedefaultmidpunct}
{\mcitedefaultendpunct}{\mcitedefaultseppunct}\relax
\EndOfBibitem
\bibitem{LHCb-DP-2012-003}
M.~Adinolfi {\em et~al.},
  \ifthenelse{\boolean{articletitles}}{\emph{{Performance of the \lhcb RICH
  detector at the LHC}},
  }{}\href{https://doi.org/10.1140/epjc/s10052-013-2431-9}{Eur.\ Phys.\ J.\
  \textbf{C73} (2013) 2431},
  \href{http://arxiv.org/abs/1211.6759}{{\normalfont\ttfamily
  arXiv:1211.6759}}\relax
\mciteBstWouldAddEndPuncttrue
\mciteSetBstMidEndSepPunct{\mcitedefaultmidpunct}
{\mcitedefaultendpunct}{\mcitedefaultseppunct}\relax
\EndOfBibitem
\bibitem{LHCb-DP-2012-002}
A.~A. Alves~Jr.\ {\em et~al.},
  \ifthenelse{\boolean{articletitles}}{\emph{{Performance of the LHCb muon
  system}}, }{}\href{https://doi.org/10.1088/1748-0221/8/02/P02022}{JINST
  \textbf{8} (2013) P02022},
  \href{http://arxiv.org/abs/1211.1346}{{\normalfont\ttfamily
  arXiv:1211.1346}}\relax
\mciteBstWouldAddEndPuncttrue
\mciteSetBstMidEndSepPunct{\mcitedefaultmidpunct}
{\mcitedefaultendpunct}{\mcitedefaultseppunct}\relax
\EndOfBibitem
\bibitem{LHCb-DP-2012-004}
R.~Aaij {\em et~al.}, \ifthenelse{\boolean{articletitles}}{\emph{{The \lhcb
  trigger and its performance in 2011}},
  }{}\href{https://doi.org/10.1088/1748-0221/8/04/P04022}{JINST \textbf{8}
  (2013) P04022}, \href{http://arxiv.org/abs/1211.3055}{{\normalfont\ttfamily
  arXiv:1211.3055}}\relax
\mciteBstWouldAddEndPuncttrue
\mciteSetBstMidEndSepPunct{\mcitedefaultmidpunct}
{\mcitedefaultendpunct}{\mcitedefaultseppunct}\relax
\EndOfBibitem
\bibitem{Sjostrand:2007gs}
T.~Sj\"{o}strand, S.~Mrenna, and P.~Skands,
  \ifthenelse{\boolean{articletitles}}{\emph{{A brief introduction to PYTHIA
  8.1}}, }{}\href{https://doi.org/10.1016/j.cpc.2008.01.036}{Comput.\ Phys.\
  Commun.\  \textbf{178} (2008) 852},
  \href{http://arxiv.org/abs/0710.3820}{{\normalfont\ttfamily
  arXiv:0710.3820}}\relax
\mciteBstWouldAddEndPuncttrue
\mciteSetBstMidEndSepPunct{\mcitedefaultmidpunct}
{\mcitedefaultendpunct}{\mcitedefaultseppunct}\relax
\EndOfBibitem
\bibitem{LHCb-PROC-2010-056}
I.~Belyaev {\em et~al.}, \ifthenelse{\boolean{articletitles}}{\emph{{Handling
  of the generation of primary events in Gauss, the LHCb simulation
  framework}}, }{}\href{https://doi.org/10.1088/1742-6596/331/3/032047}{J.\
  Phys.\ Conf.\ Ser.\  \textbf{331} (2011) 032047}\relax
\mciteBstWouldAddEndPuncttrue
\mciteSetBstMidEndSepPunct{\mcitedefaultmidpunct}
{\mcitedefaultendpunct}{\mcitedefaultseppunct}\relax
\EndOfBibitem
\bibitem{Chang:2005hq}
C.-H. Chang, J.-X. Wang, and X.-G. Wu,
  \ifthenelse{\boolean{articletitles}}{\emph{{BCVEGPY2.0: A upgrade version of
  the generator BCVEGPY with an addendum about hadroproduction of the $P$-wave
  $B_c$ states}}, }{}\href{https://doi.org/10.1016/j.cpc.2005.09.008}{Comput.\
  Phys.\ Commun.\  \textbf{174} (2006) 241},
  \href{http://arxiv.org/abs/hep-ph/0504017}{{\normalfont\ttfamily
  arXiv:hep-ph/0504017}}\relax
\mciteBstWouldAddEndPuncttrue
\mciteSetBstMidEndSepPunct{\mcitedefaultmidpunct}
{\mcitedefaultendpunct}{\mcitedefaultseppunct}\relax
\EndOfBibitem
\bibitem{Lange:2001uf}
D.~J. Lange, \ifthenelse{\boolean{articletitles}}{\emph{{The EvtGen particle
  decay simulation package}},
  }{}\href{https://doi.org/10.1016/S0168-9002(01)00089-4}{Nucl.\ Instrum.\
  Meth.\  \textbf{A462} (2001) 152}\relax
\mciteBstWouldAddEndPuncttrue
\mciteSetBstMidEndSepPunct{\mcitedefaultmidpunct}
{\mcitedefaultendpunct}{\mcitedefaultseppunct}\relax
\EndOfBibitem
\bibitem{Golonka:2005pn}
P.~Golonka and Z.~Was, \ifthenelse{\boolean{articletitles}}{\emph{{PHOTOS Monte
  Carlo: A precision tool for QED corrections in $Z$ and $W$ decays}},
  }{}\href{https://doi.org/10.1140/epjc/s2005-02396-4}{Eur.\ Phys.\ J.\
  \textbf{C45} (2006) 97},
  \href{http://arxiv.org/abs/hep-ph/0506026}{{\normalfont\ttfamily
  arXiv:hep-ph/0506026}}\relax
\mciteBstWouldAddEndPuncttrue
\mciteSetBstMidEndSepPunct{\mcitedefaultmidpunct}
{\mcitedefaultendpunct}{\mcitedefaultseppunct}\relax
\EndOfBibitem
\bibitem{Allison:2006ve}
Geant4 collaboration, J.~Allison {\em et~al.},
  \ifthenelse{\boolean{articletitles}}{\emph{{Geant4 developments and
  applications}}, }{}\href{https://doi.org/10.1109/TNS.2006.869826}{IEEE
  Trans.\ Nucl.\ Sci.\  \textbf{53} (2006) 270}\relax
\mciteBstWouldAddEndPuncttrue
\mciteSetBstMidEndSepPunct{\mcitedefaultmidpunct}
{\mcitedefaultendpunct}{\mcitedefaultseppunct}\relax
\EndOfBibitem
\bibitem{Agostinelli:2002hh}
Geant4 collaboration, S.~Agostinelli {\em et~al.},
  \ifthenelse{\boolean{articletitles}}{\emph{{Geant4: A simulation toolkit}},
  }{}\href{https://doi.org/10.1016/S0168-9002(03)01368-8}{Nucl.\ Instrum.\
  Meth.\  \textbf{A506} (2003) 250}\relax
\mciteBstWouldAddEndPuncttrue
\mciteSetBstMidEndSepPunct{\mcitedefaultmidpunct}
{\mcitedefaultendpunct}{\mcitedefaultseppunct}\relax
\EndOfBibitem
\bibitem{LHCb-PROC-2011-006}
M.~Clemencic {\em et~al.}, \ifthenelse{\boolean{articletitles}}{\emph{{The
  \lhcb simulation application, Gauss: Design, evolution and experience}},
  }{}\href{https://doi.org/10.1088/1742-6596/331/3/032023}{J.\ Phys.\ Conf.\
  Ser.\  \textbf{331} (2011) 032023}\relax
\mciteBstWouldAddEndPuncttrue
\mciteSetBstMidEndSepPunct{\mcitedefaultmidpunct}
{\mcitedefaultendpunct}{\mcitedefaultseppunct}\relax
\EndOfBibitem
\bibitem{Breiman:1984jka}
L.~Breiman, J.~Friedman, R.~A. Olshen, and C.~J. Stone, {\em {Classification
  and regression trees}}, Chapman and Hall/CRC, 1984\relax
\mciteBstWouldAddEndPuncttrue
\mciteSetBstMidEndSepPunct{\mcitedefaultmidpunct}
{\mcitedefaultendpunct}{\mcitedefaultseppunct}\relax
\EndOfBibitem
\bibitem{Freund:1997xna}
Y.~Freund and R.~E. Schapire, \ifthenelse{\boolean{articletitles}}{\emph{{A
  decision-theoretic generalization of on-line learning and an application to
  boosting}}, }{}\href{https://doi.org/10.1006/jcss.1997.1504}{J.\ Comput.\
  Syst.\ Sci.\  \textbf{55} (1997) 119}\relax
\mciteBstWouldAddEndPuncttrue
\mciteSetBstMidEndSepPunct{\mcitedefaultmidpunct}
{\mcitedefaultendpunct}{\mcitedefaultseppunct}\relax
\EndOfBibitem
\bibitem{friedman2001}
J.~H. Friedman, \ifthenelse{\boolean{articletitles}}{\emph{Greedy function
  approximation: A gradient boosting machine.},
  }{}\href{https://doi.org/10.1214/aos/1013203451}{Ann.\ Statist.\  \textbf{29}
  (2001) 1189}\relax
\mciteBstWouldAddEndPuncttrue
\mciteSetBstMidEndSepPunct{\mcitedefaultmidpunct}
{\mcitedefaultendpunct}{\mcitedefaultseppunct}\relax
\EndOfBibitem
\bibitem{TMVA4}
A.~Hoecker {\em et~al.}, \ifthenelse{\boolean{articletitles}}{\emph{{TMVA 4 ---
  Toolkit for Multivariate Data Analysis with ROOT. Users Guide.}},
  }{}\href{http://arxiv.org/abs/physics/0703039}{{\normalfont\ttfamily
  arXiv:physics/0703039}}\relax
\mciteBstWouldAddEndPuncttrue
\mciteSetBstMidEndSepPunct{\mcitedefaultmidpunct}
{\mcitedefaultendpunct}{\mcitedefaultseppunct}\relax
\EndOfBibitem
\bibitem{Skwarnicki:1986xj}
T.~Skwarnicki, {\em {A study of the radiative cascade transitions between the
  Upsilon-prime and Upsilon resonances}}, PhD thesis, Institute of Nuclear
  Physics, Krakow, 1986,
  {\href{http://inspirehep.net/record/230779/}{DESY-F31-86-02}}\relax
\mciteBstWouldAddEndPuncttrue
\mciteSetBstMidEndSepPunct{\mcitedefaultmidpunct}
{\mcitedefaultendpunct}{\mcitedefaultseppunct}\relax
\EndOfBibitem
\bibitem{Hulsbergen:2005pu}
W.~D. Hulsbergen, \ifthenelse{\boolean{articletitles}}{\emph{{Decay chain
  fitting with a Kalman filter}},
  }{}\href{https://doi.org/10.1016/j.nima.2005.06.078}{Nucl.\ Instrum.\ Meth.\
  \textbf{A552} (2005) 566},
  \href{http://arxiv.org/abs/physics/0503191}{{\normalfont\ttfamily
  arXiv:physics/0503191}}\relax
\mciteBstWouldAddEndPuncttrue
\mciteSetBstMidEndSepPunct{\mcitedefaultmidpunct}
{\mcitedefaultendpunct}{\mcitedefaultseppunct}\relax
\EndOfBibitem
\bibitem{LHCb-PAPER-2018-046}
LHCb collaboration, R.~Aaij {\em et~al.},
  \ifthenelse{\boolean{articletitles}}{\emph{{Observation of
  \mbox{\decay{\BdorBs}{\jpsi \proton\antiproton}} decays and precision
  measurements of the \BdorBs masses}},
  }{}\href{https://doi.org/10.1103/PhysRevLett.122.191804}{Phys.\ Rev.\ Lett.\
  \textbf{122} (2019) 191804},
  \href{http://arxiv.org/abs/1902.05588}{{\normalfont\ttfamily
  arXiv:1902.05588}}\relax
\mciteBstWouldAddEndPuncttrue
\mciteSetBstMidEndSepPunct{\mcitedefaultmidpunct}
{\mcitedefaultendpunct}{\mcitedefaultseppunct}\relax
\EndOfBibitem
\bibitem{LHCb-PAPER-2018-018}
LHCb collaboration, R.~Aaij {\em et~al.},
  \ifthenelse{\boolean{articletitles}}{\emph{{Observation of the decay
  \mbox{\decay{\Bsb}{\chi_{c2}\Kp\Km}}}},
  }{}\href{https://doi.org/10.1007/JHEP08(2018)191}{JHEP \textbf{08} (2018)
  191}, \href{http://arxiv.org/abs/1806.10576}{{\normalfont\ttfamily
  arXiv:1806.10576}}\relax
\mciteBstWouldAddEndPuncttrue
\mciteSetBstMidEndSepPunct{\mcitedefaultmidpunct}
{\mcitedefaultendpunct}{\mcitedefaultseppunct}\relax
\EndOfBibitem
\bibitem{LHCb-PAPER-2015-033}
LHCb collaboration, R.~Aaij {\em et~al.},
  \ifthenelse{\boolean{articletitles}}{\emph{{Observation of the
  \mbox{\decay{\Bs}{\jpsi\phi\phi}} decay}},
  }{}\href{https://doi.org/10.1007/JHEP03(2016)040}{JHEP \textbf{03} (2016)
  040}, \href{http://arxiv.org/abs/1601.05284}{{\normalfont\ttfamily
  arXiv:1601.05284}}\relax
\mciteBstWouldAddEndPuncttrue
\mciteSetBstMidEndSepPunct{\mcitedefaultmidpunct}
{\mcitedefaultendpunct}{\mcitedefaultseppunct}\relax
\EndOfBibitem
\bibitem{LHCb-PAPER-2011-035}
LHCb collaboration, R.~Aaij {\em et~al.},
  \ifthenelse{\boolean{articletitles}}{\emph{{Measurement of \bquark-hadron
  masses}}, }{}\href{https://doi.org/10.1016/j.physletb.2012.01.058}{Phys.\
  Lett.\  \textbf{B708} (2012) 241},
  \href{http://arxiv.org/abs/1112.4896}{{\normalfont\ttfamily
  arXiv:1112.4896}}\relax
\mciteBstWouldAddEndPuncttrue
\mciteSetBstMidEndSepPunct{\mcitedefaultmidpunct}
{\mcitedefaultendpunct}{\mcitedefaultseppunct}\relax
\EndOfBibitem
\bibitem{LHCb-PAPER-2017-018}
LHCb collaboration, R.~Aaij {\em et~al.},
  \ifthenelse{\boolean{articletitles}}{\emph{{Observation of the doubly charmed
  baryon \Xiccpp}},
  }{}\href{https://doi.org/10.1103/PhysRevLett.119.112001}{Phys.\ Rev.\ Lett.\
  \textbf{119} (2017) 112001},
  \href{http://arxiv.org/abs/1707.01621}{{\normalfont\ttfamily
  arXiv:1707.01621}}\relax
\mciteBstWouldAddEndPuncttrue
\mciteSetBstMidEndSepPunct{\mcitedefaultmidpunct}
{\mcitedefaultendpunct}{\mcitedefaultseppunct}\relax
\EndOfBibitem
\bibitem{Santos:2013gra}
D.~Mart{\'\i}nez~Santos and F.~Dupertuis,
  \ifthenelse{\boolean{articletitles}}{\emph{{Mass distributions marginalized
  over per-event errors}},
  }{}\href{https://doi.org/10.1016/j.nima.2014.06.081}{Nucl.\ Instrum.\ Meth.\
  \textbf{A764} (2014) 150},
  \href{http://arxiv.org/abs/1312.5000}{{\normalfont\ttfamily
  arXiv:1312.5000}}\relax
\mciteBstWouldAddEndPuncttrue
\mciteSetBstMidEndSepPunct{\mcitedefaultmidpunct}
{\mcitedefaultendpunct}{\mcitedefaultseppunct}\relax
\EndOfBibitem
\bibitem{Lyons:1988rp}
L.~Lyons, D.~Gibaut, and P.~Clifford,
  \ifthenelse{\boolean{articletitles}}{\emph{{How to combine correlated
  estimates of a single physical quantity}},
  }{}\href{https://doi.org/10.1016/0168-9002(88)90018-6}{Nucl.\ Instrum.\
  Meth.\  \textbf{A270} (1988) 110}\relax
\mciteBstWouldAddEndPuncttrue
\mciteSetBstMidEndSepPunct{\mcitedefaultmidpunct}
{\mcitedefaultendpunct}{\mcitedefaultseppunct}\relax
\EndOfBibitem
\bibitem{Valassi:2003mu}
A.~Valassi, \ifthenelse{\boolean{articletitles}}{\emph{{Combining correlated
  measurements of several different physical quantities}},
  }{}\href{https://doi.org/10.1016/S0168-9002(03)00329-2}{Nucl.\ Instrum.\
  Meth.\  \textbf{A500} (2003) 391}\relax
\mciteBstWouldAddEndPuncttrue
\mciteSetBstMidEndSepPunct{\mcitedefaultmidpunct}
{\mcitedefaultendpunct}{\mcitedefaultseppunct}\relax
\EndOfBibitem
\bibitem{Nisius:2020jmf}
R.~Nisius, \ifthenelse{\boolean{articletitles}}{\emph{{BLUE: combining
  correlated estimates of physics observables within ROOT using the Best Linear
  Unbiased Estimate method}},
  }{}\href{https://doi.org/10.1016/j.softx.2020.100468}{SoftwareX \textbf{11}
  (2020) 100468}, \href{http://arxiv.org/abs/2001.10310}{{\normalfont\ttfamily
  arXiv:2001.10310}}\relax
\mciteBstWouldAddEndPuncttrue
\mciteSetBstMidEndSepPunct{\mcitedefaultmidpunct}
{\mcitedefaultendpunct}{\mcitedefaultseppunct}\relax
\EndOfBibitem
\end{mcitethebibliography}

\newpage
\centerline
{\large\bf LHCb collaboration}
\begin
{flushleft}
\small
R.~Aaij$^{31}$,
C.~Abell{\'a}n~Beteta$^{49}$,
T.~Ackernley$^{59}$,
B.~Adeva$^{45}$,
M.~Adinolfi$^{53}$,
H.~Afsharnia$^{9}$,
C.A.~Aidala$^{81}$,
S.~Aiola$^{25}$,
Z.~Ajaltouni$^{9}$,
S.~Akar$^{66}$
% ...

{\footnotesize \it

% ...
$ ^{9}$Universit{\'e} Clermont Auvergne, CNRS/IN2P3, LPC, Clermont-Ferrand, France\\
% ...
$ ^{25}$INFN Sezione di Milano, Milano, Italy\\
% ...
$ ^{31}$Nikhef National Institute for Subatomic Physics, Amsterdam, Netherlands\\
% ...
$ ^{45}$Instituto Galego de F{\'\i}sica de Altas Enerx{\'\i}as (IGFAE), Universidade de Santiago de Compostela, Santiago de Compostela, Spain\\
% ...
$ ^{49}$Physik-Institut, Universit{\"a}t Z{\"u}rich, Z{\"u}rich, Switzerland\\
% ...
$ ^{53}$H.H. Wills Physics Laboratory, University of Bristol, Bristol, United Kingdom\\
% ...
$ ^{59}$Oliver Lodge Laboratory, University of Liverpool, Liverpool, United Kingdom\\
% ...
$ ^{66}$Los Alamos National Laboratory (LANL), Los Alamos, United States\\
$ ^{67}$Syracuse University, Syracuse, NY, United States\\
% ...
$ ^{81}$University of Michigan, Ann Arbor, United States, associated to $^{67}$\\
\bigskip
% ...
\medskip
}
\end{flushleft}

\end{document}